\documentclass[aps,prb,twocolumn,reprint,floatfix,nofootinbib]{revtex4-2}
\usepackage{graphicx}
\usepackage{amsmath,amssymb,mathrsfs,amsthm}
\usepackage[pdftex, colorlinks=true, linkcolor=blue, citecolor=blue, urlcolor=blue]{hyperref}
\usepackage{bbold}
\usepackage{times}
\usepackage{mathtools}
\usepackage{braket}

\usepackage{lipsum}
\usepackage{svg}
\usepackage[normalem]{ulem}
\usepackage{siunitx}
\usepackage{color}

\usepackage{float}
\newcommand{\stkout}[1]{\ifmmode\text{\sout{\ensuremath{#1}}}\else\sout{#1}\fi}

\begin{document}

\title{Mirror-induced effects in cavity polaritonics: Influence on edge states}

\author{Thomas F.\ Allard}
\affiliation{Universit\'e de Strasbourg, CNRS, Institut de Physique et Chimie des Mat\'eriaux de Strasbourg, UMR 7504, F-67000 Strasbourg, France}
\author{Guillaume Weick}
\affiliation{Universit\'e de Strasbourg, CNRS, Institut de Physique et Chimie des Mat\'eriaux de Strasbourg, UMR 7504, F-67000 Strasbourg, France}

\begin{abstract}

Optical cavities are widely used to induce strong light-matter coupling and thereby enable the presence of polaritons.
While polaritons are at the source of most of the observed physics, the mirrors forming the cavity may also themselves be responsible for a number of phenomena, independently of the strong light-matter coupling regime.
Here we use a toy model of a chain of dipolar emitters coupled to a cuboidal cavity. We unveil several effects originating solely from the boundary conditions imposed by the cavity mirrors, that are dominant when the distances of the emitters to the cavity walls are of the order of the interdipole separation. In particular, 
we show that mirrors in the direction transverse to the chain may act as effective defects, leading to the emergence of Tamm edge states.
Considering a topological chain, we demonstrate that such transverse mirrors may also protect edge states against the effects of the strong light-matter coupling.
Finally, we find that mirrors parallel to the chain, by the image charges they involve, induce topological phase transitions even in the case of highly off-resonant photons.

\end{abstract}

\maketitle

\section{Introduction}
\label{sec:Introduction}

\begin{figure*}[tb]
 \includegraphics[width=.89\linewidth]{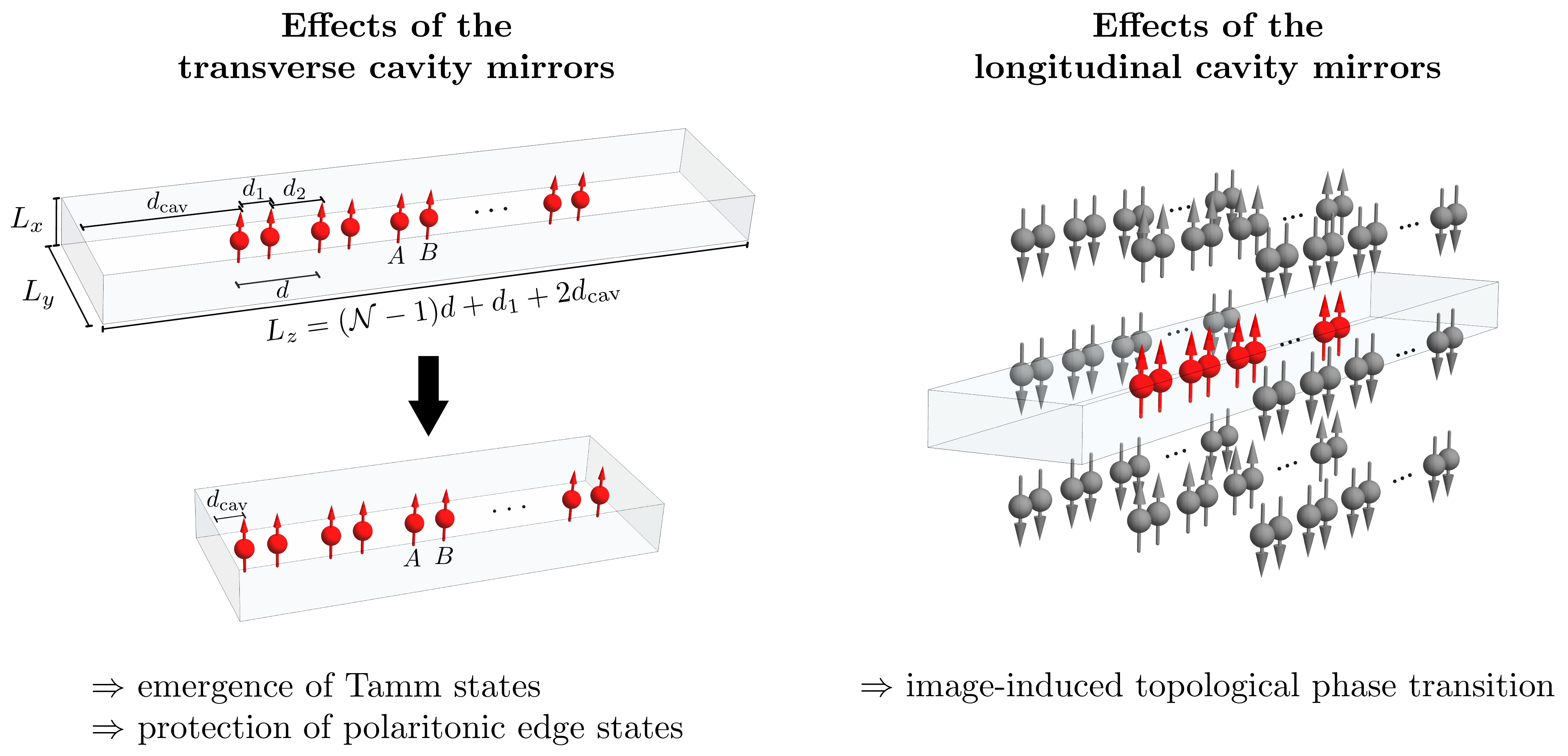}
 \caption{Sketch summarizing the system under study in this work as well as the mirror-induced effects we unveil.
 A dimerized chain of $2\mathcal{N}$ emitters is embedded in a closed cuboidal cavity with perfect metallic mirrors in all $x$, $y$, and $z$ directions.
 The emitters, belonging either to the $A$ or $B$ sublattice, are all polarized along the $x$ axis, have all the same bare frequency $\omega_0$, and are separated from each other by the alternating distances $d_1$ and $d_2$.
 The first and last emitters of the chain are placed at a distance $d_\mathrm{cav}$ from the cavity walls in the $z$ direction, which we call transverse cavity mirrors.
 In Sec.~\ref{sec:transverse}, we investigate the effect of these transverse cavity mirrors and unveil the emergence of Tamm states and the protection of edge states of topological origin as the distance $d_\mathrm{cav}$ is reduced.
 In Sec.~\ref{sec:Images}, we explore the specific impact of the longitudinal cavity mirrors, which generate image dipoles (represented in gray) that renormalize the dipolar degrees of freedom and may induce topological phase transitions.}
 \label{fig:Sketch}
\end{figure*}

The past decades have witnessed the emergence of strongly coupled light-matter systems, designed to take advantage of the special characteristics of hybrid light and matter excitations called polaritons \cite{EbbesenReview2021}.
A variety of experimental platforms have been shown to host such polaritons, from organic semiconductors \cite{Lidzey1998} to plasmonic nanoparticles \cite{Baranov2020}, macroscopic microwave resonators \cite{Zhao2024_arXiv}, or cold atoms \cite{Sauerwein2023,Baghdad2023}, among many others \cite{Baranov2018,Basov2020}.
While some of these systems rely on the extreme natural coupling of matter degrees of freedom to the vacuum electromagnetic field \cite{Mueller2020,Thomas2021}, most of them require the use of optical cavities in order to enhance the light-matter interaction.
Many exciting phenomena induced by polaritons have been studied, such as long-range charge and energy transfer \cite{Takazawa2010,Orgiu2015}, protection against disorder \cite{Sauerwein2023,Baghdad2023}, the alteration of chemical reactions \cite{Hutchison2012,Ribeiro2018}, or an exotic interplay with topological matter, such as in the integer \cite{Appugliese2022} or fractional quantum Hall effect \cite{Enkner2024}.
However, as the systems under study may be very complex, identifying precisely effects resulting strictly from the polaritonic nature of the excitations may be difficult, leading to recent discussions on the possibility of cavity-induced nonpolaritonic effects \cite{Thomas2024_JPCL,Thomas2024_AdvMat, Schwartz_comment} or optical artifacts due to inhomogeneities of the cavity walls \cite{Piejko2023, Piejko2024}. 

Motivated by these recent debates, in this work we aim at studying the direct effects that cavity mirrors may induce on matter degrees of freedom, independently of the strong light-matter coupling regime.
Indeed, mirrors, through the boundary conditions that they impose on the electromagnetic field, can be responsible for many phenomena. 
Various boundary conditions have been used in the theoretical literature treating polaritons in optical cavities, the cavity being sometimes considered finite with hard wall boundaries \cite{Downing2019, Allard2022} or periodic \cite{Michetti_2005,Tichauer2021,Ribeiro2022,Ribeiro2023,Engelhardt2023,Allard2023}.

Here we use a simple model recently realized experimentally \cite{Zhao2024_arXiv} of a chain of ideal dipolar emitters embedded in a finite cuboidal multimode cavity made of perfect metallic mirrors (see Fig.~\ref{fig:Sketch}) to unveil specific effects that originate solely from the boundary conditions that such a cavity imposes on the electromagnetic field.
These effects are particularly prominent when the distance of the emitters to the cavity walls are of the order of their typical separation, as is the case, e.g., in the experiment of Ref.~\cite{Zhao2024_arXiv}. 

Specifically, we investigate two features often disregarded in the existing literature.
First, how the considered boundary conditions of the cavity may affect the properties of the system, through a precise study of the influence of the distance between the first and last emitters of the chain and the cavity walls.
Second, how image dipoles originating from the cavity metallic walls may affect the system.
By developing an effective Hamiltonian description of the system, we study in detail the influence of the distance between the emitters and the cavity mirrors and reveal how mirrors renormalize both the bare frequencies and the dipole-dipole coupling, independently of the strong-coupling regime.

The system under consideration in this work along with the main results we obtain are illustrated in Fig.~\ref{fig:Sketch}.
We demonstrate that mirrors in the direction transverse to the dipolar chain may act as effective defects for the matter degrees of freedom, leading to the presence of Tamm edge states at the two ends of the chain \cite{Tamm1932,Ohno1990}.
While these cavity-induced Tamm edge states, first uncovered in Ref.~\cite{Downing2021}, were interpreted as polaritonic effects, here we show how they result exclusively from the boundary conditions imposed by the mirrors.
We then study a dimerized topological dipolar chain, which, once embedded in a cavity, leads to the polaritonic Su-Schrieffer-Heeger (SSH) model introduced in Ref.~\cite{Downing2019} and subsequently investigated in detail in  Ref.~\cite{Allard2023}.
By tuning the distance between the chain and the transverse mirrors while being in the topologically nontrivial phase of the system, we show how these mirrors can assist localization and crucially protect edge states against the effects of the strong light-matter coupling.
Finally, we study the influence of image dipoles generated by mirrors parallel to the direction of the chain.
In particular, we unveil a rich phase diagram with topological phase transitions induced by such image dipoles, occurring for cavity walls close enough to the emitters and highly off-resonant photons.

The paper is organized as follows. Section~\ref{sec:Model} presents our model of dipolar emitters coupled to a finite cuboidal cavity. 
In Sec.~\ref{sec:Tamm}, we discuss how transverse mirrors can act as defects and lead to Tamm (nontopological) edge states. In Sec.~\ref{sec:Edge states}, we show the protection effect that mirrors can have on dimerized-induced edge states against polaritonic effects, while in Sec.~\ref{sec:Images} we unveil topological phase transitions induced by image dipoles.
Finally, in Sec.~\ref{sec:Conclusion} we draw conclusions on our work and outline some perspectives on the specific effects of mirrors in polaritonic systems.

\section{Dipolar emitters coupled to a cuboidal cavity}
\label{sec:Model}

To investigate the specific effects of cavity mirrors on polaritonic systems, we consider a simple model of a chain of dipolar emitters coupled to a cuboidal metallic cavity, as sketched in Fig.~\ref{fig:Sketch}.
The generic emitters considered in this work are modeled as subwavelength point dipoles without any internal degrees of freedom, so that they behave as classical oscillating dipoles, and may represent diverse dipolar physical systems that are governed by classical electrodynamics.
Typical examples of experimental platforms modeled by such classical dipolar emitters are subwavelength plasmonic \cite{Mueller2020}, dielectric \cite{Slobozhanyuk2015} or SiC \cite{Wang2018b} nanoparticles, macroscopic helical microwave antennas \cite{Mann2018,Mann2020,Zhao2024_arXiv}, cold Rydberg atoms \cite{Browaeys2016} (or any other two-level emitters in the single excitation manifold \cite{Asenjo-Garcia2019}), magnonic microspheres \cite{Rameshti2022}, or semiconductor excitons \cite{Yuen-Zhou2016}.

Since we will in particular investigate topological effects in the following sections, we consider a possibly dimerized chain with emitters belonging either to the $A$ or $B$ sublattice, a geometry analogous to the one of the celebrated SSH model of one-dimensional topological physics \cite{Su1979,Asboth2016}.
Such a polaritonic model has been originally developed in Ref.~\cite{Downing2019}, and later further studied in recent works \cite{Downing2021,Allard2022,Allard2023}.
Importantly, the model includes the quasistatic coupling between the emitters as well as the coupling to many cavity modes, hence going beyond the widely used Tavis-Cummings model  \cite{Tavis1968}.
We note that the necessity of taking into account  multiple photonic modes to properly address polaritonic properties has been recently highlighted in the literature, notably in the context of molecular polaritons \cite{Allard2022,Tichauer2021,Ribeiro2022,Ribeiro2023,Engelhardt2023,Mandal2023}.

Crucially, in the present work, we go beyond the model discussed in Refs.~\cite{Downing2019,Downing2021,Allard2022,Allard2023} as we also take into account specific effects induced by the metallic cavity such as the renormalization of the dipolar degrees of freedom due to image dipoles, as well as the influence of the distance $d_\mathrm{cav}$ between the chain of emitters and the transverse cavity walls  (see Fig.~\ref{fig:Sketch}).
This allows us to decipher between polaritonic and mirror-induced effects in the system.

\subsection{Polaritonic Hamiltonian}

The model Hamiltonian can be separated in three parts and reads as follows:
\begin{equation}
 H_\mathrm{pol} = H^\mathrm{im}_\mathrm{dp} + H_\mathrm{ph} + H_\mathrm{dp\textrm{-}ph}.
\label{eq:total Hamiltonian}
\end{equation}
The first part, $H^\mathrm{im}_\mathrm{dp}$, contains the dipolar degrees of freedom.
Each dipolar emitter belongs to a unit cell labeled by $m$ ($m=1, \ldots, \mathcal{N}$) and is modeled through a single dynamical degree of freedom, its displacement field $\mathbf{h}_m^s$ oriented along the $x$ axis, with $s=A,B$ the sublattice index.
The emitters all oscillate at the same bare frequency $\omega_0$, and their internal properties such as their (effective) masses and charges are encapsulated in their typical dipole length scale $a$.
The dimerized chain, sketched in Fig.~\ref{fig:Sketch}, consists of emitters separated by alternating distances $d_1$ and $d_2$, with $d=d_1+d_2$ the interdimer distance.
To characterize the dimerization of the chain, we define the parameter $\epsilon = (d_1 - d_2)/d$, so that $\epsilon=0$ corresponds to a regular chain.

The quasistatic Hamiltonian $H^\mathrm{im}_\mathrm{dp}$ includes the bare onsite energies $\omega_0$ of the emitters as well as the one over the distance cubed Coulomb dipole-dipole coupling between them.
Both are renormalized by image dipoles originating from the boundary conditions imposed by the cavity, the onsite energies being redshifted and the quasistatic coupling being exponentially suppressed as the metallic walls of the cavity get closer to the emitters.
Adopting the rotating wave approximation (RWA), where the number of excitations is conserved, such a dipolar Hamiltonian reads (see Appendix~\ref{sec:Appendix images} for details)
\begin{align}
    H_\mathrm{dp}^\mathrm{im} =&\; \sum_{m=1}^\mathcal{N} 
    \left( \hbar \omega_m^{\mathrm{im},A}  a_{m}^{\dagger} a_{m}^{\phantom{\dagger}} + \hbar \omega_m^{\mathrm{im},B} b_{m}^{\dagger} b_{m}^{\phantom{\dagger}}\right)
    \nonumber \\ 
    &+ \frac{\hbar\Omega}{2}\sum_{\substack{m,m'=1\\(m\neq m')}}^{\mathcal{N}} \left[ f_{m,m'}^{\mathrm{im},A} \left( a_{m}^{\dagger} a_{m^\prime}^{\phantom{\dagger}} + \mathrm{H.c.} \right) \right. \nonumber \\
    & \left. +\; f_{m,m'}^{\mathrm{im},B} \left( b_{m}^{\dagger} b_{m^\prime}^{\phantom{\dagger}} + \mathrm{H.c.} \right) \right]
    \nonumber \\
    &+ \hbar\Omega\sum_{m,m'=1}^{\mathcal{N}} g_{m,m'}^\mathrm{im} \left( a_{m}^{\dagger} b_{m^\prime}^{\phantom{\dagger}} + \mathrm{H.c.} \right),
\label{eq:H_dp with images}
\end{align}
where $\Omega=(\omega_0/2)(a/d)^3$ is the quasistatic dipolar coupling strength. Here the bosonic operator $a_m$/$b_m$ ($a_m^\dagger$/$b_m^\dagger$) annihilates (creates) a dipolar excitation on the site $A$/$B$ of the unit cell $m$.\footnote{Note that for the single-particle, quadratic, and spinless system under consideration, the specific quantum statistics of the excitations is inconsequential.}
The Hamiltonian \eqref{eq:H_dp with images} can be interpreted as a one-particle tight-binding Hamiltonian, with the quantities $\omega_m^{\mathrm{im}, A}$, $\omega_m^{\mathrm{im},B}$, $f_{m,m'}^{\mathrm{im}, A}$, $f_{m,m'}^{\mathrm{im},B}$, and $g_{m,m'}^\mathrm{im}$ denoting, respectively, the image-renormalized onsite frequencies of the emitters, and the intrasublattice ($A \leftrightarrow A$ and $B \leftrightarrow B$) and intersublattice ($A \leftrightarrow B$) couplings between the emitters.
The detailed expressions of these five quantities are given in Appendix~\ref{sec:Appendix images} [cf.\ Eq.~\eqref{eq:app:image renormalized quantities}].
While in most cases the renormalizations induced by image dipoles are only of quantitative incidence on the prevailing physics, we shall see in Sec.~\ref{sec:Images} that for cavity walls close enough to the emitters their effect may be preponderant.

The second and third parts of the Hamiltonian \eqref{eq:total Hamiltonian} denote the photonic degrees of freedom of the cavity and their coupling to the emitters.
From the quantization of the electromagnetic field in the cuboidal cavity, the photon wave vector is quantized as $\mathbf{k} = (\pi n_x/L_x , \pi n_y/L_y , \pi n_z/L_z)$, with $(n_x, n_y, n_z) \in \mathbb{N}^3 \backslash (0,0,0)$ and $L_x$, $L_y$, and $L_z$ respectively the height, width, and length of the cavity (see Fig.~\ref{fig:Sketch}).
As discussed in detail in Refs.~\cite{Downing2019,Downing2021}, by assuming a cavity geometry with a width $L_y$ larger than its height $L_x$ and emitters placed at the center of the cavity, one can approximate the light-matter coupling to only the lowest photonic band $(n_x,n_y,n_z) = (0,1,n_z)$, the others being at much higher energies.
By fixing the aspect ratio of the cavity to $L_y/L_x=3$ and the typical dipole length scale $a$ such that $\omega_0a/c=0.1$ (ensuring the point dipole approximation), such a single cavity band approximation is valid as long as the cavity height $L_x$ is approximately in the range $[2a,15a]$, for which the lowest band is the only one at resonance with the dipolar emitters.

The above considerations allow us to write the photonic Hamiltonian $H_\mathrm{ph}$ as
\begin{equation}
	H_\mathrm{ph} = \sum_{n_z=1}^{N_z} \hbar \omega^{\mathrm{ph}}_{n_z} c^{\dagger}_{n_z} c^{\phantom\dagger}_{n_z},
\label{eq:H_ph}
\end{equation}
with the photonic dispersion
\begin{equation}
	\omega^\mathrm{ph}_{n_z} = c \sqrt{ \left(\frac{\pi}{L_y}\right)^2 + \left(\frac{\pi n_z}{L_z}\right)^2 },
\label{eq:photonic dispersion}
\end{equation}
with $c$ being the speed of light in vacuum and where $c_{n_z}$ ($c_{n_z}^\dagger$) is the bosonic annihilation (creation) operator of a photonic excitation with longitudinal wave vector $k_z = \pi n_z /L_z$.\footnote{Note that the number $N_z$ of photonic modes entering the Hamiltonian \eqref{eq:H_ph} must, for computational purposes, be fixed to a given value.
In the remaining of the paper, we set $N_z = \left \lfloor 20L_z/d \right \rfloor$, so that the maximal longitudinal photon wave number $\mathrm{max}(k_z) = 20\pi/d$, a value that we have verified to be large enough for the convergence of our numerical results.}
Within the RWA, the single-band light-matter coupling Hamiltonian $H_\mathrm{dp\textrm{-}ph}$ in Eq.~\eqref{eq:total Hamiltonian} then reads
\begin{align}
	H_\mathrm{dp\textrm{-}ph} =&\; \mathrm{i}\hbar \sum_{m=1}^\mathcal{N} \sum_{n_z=1}^{N_z}  \left[ \xi^{A}_{m n_z} \left( a^\dagger_m c^{\phantom\dagger}_{n_z} - a^{\phantom\dagger}_m c^\dagger_{n_z} \right) \right. \nonumber \\
    & \left. +\; \xi^{B}_{m n_z} \left( b^\dagger_m c^{\phantom\dagger}_{n_z} - b^{\phantom\dagger}_m c^\dagger_{n_z} \right) \right].
\label{eq:H_dp-ph}
\end{align}
Here the mode-dependent light-matter coupling functions acting on the $A$ and $B$ sublattices are
\begin{equation}
	\xi^{A(B)}_{mn_z} = \omega_0\sqrt{ \frac{4\pi a^3\omega_0}{L_x L_y L_z \omega^{\mathrm{ph}}_{n_z}} }\sin\left(\frac{\pi n_z}{L_z}z^{A(B)}_m \right),
\label{eq:light-matter coupling functions A and B}
\end{equation}
which depend on the $z$ coordinate of an emitter, given by 
\begin{subequations}
\label{eq:zsm}
\begin{align}
z^A_m &= (m-1)d + d_\mathrm{cav},\\
z^B_m &= (m-1)d + d_1 + d_\mathrm{cav},
\end{align}
\end{subequations}
where the origin is located at the left transverse cavity wall.
It is noteworthy that having set the aspect ratio, for a given chain with fixed number of dimers $\mathcal{N}$ and fixed distance to the transverse cavity walls $d_\mathrm{cav}$, the only parameter left that handles the photonic degrees of freedom and their coupling to the dipolar emitters is the cavity height $L_x/a$ (in units of the dipole length scale $a$).

\subsection{Polaritonic dispersion}

In order to first gain some insight into the present model of emitters in a multimode cavity, we here briefly discuss its dispersion relation and eigenstates at the thermodynamic limit, therefore considering an infinite waveguide cavity.
This analysis has already been performed in the absence of image dipoles in Ref.~\cite{Downing2019}, while the results which we discuss below do include them.

The Hamiltonian \eqref{eq:total Hamiltonian} can be rewritten in Fourier space, considering an infinite and periodic chain of $\mathcal{N}\to\infty$ pairs of emitters with lattice constant $d$, along with an infinitely extended cavity of length $L_z \sim \mathcal{N}d \to \infty$.
The translational invariance then allows us to assume periodic boundary conditions for the electromagnetic field within the cavity in the $z$ direction.
It leads the longitudinal photon wave number $k_z$ to be conserved with the newly defined quasimomentum $q$ of the infinite periodic chain of emitters, which belongs to the first Brillouin zone $[-\pi/d,+\pi/d]$.

The above-mentioned procedure, detailed in Appendix~\ref{sec:Appendix Fourier transformation}, allows us to rewrite Eq.~\eqref{eq:total Hamiltonian} as a three-band Hamiltonian, which can then be diagonalized analytically as
\begin{equation}
    H_\mathrm{pol} = \sum_q \sum_j \hbar\omega^\mathrm{pol}_{qj} \gamma_{qj}^\dagger \gamma^{\phantom\dagger}_{qj}.
\label{eq:total Hamiltonian fourier}
\end{equation}
The analytical expressions of the polaritonic dispersion $\omega^\mathrm{pol}_{qj}$, with the index $j=1,2,3$ labeling, respectively, the upper (UP), middle (MP), and lower (LP) polariton, and of the polaritonic eigenstates $\gamma_{qj}$, which are a linear combination of dipolar and photonic degrees of freedom, are given in Appendix~\ref{sec:Appendix Fourier transformation} [cf.\ Eqs.~\eqref{eq:app:full polaritonic dispersion} and \eqref{eq:app:full polaritonic Hamiltonian Hopfield operator}, respectively].

\begin{figure}[tb]
\includegraphics[width=\linewidth]{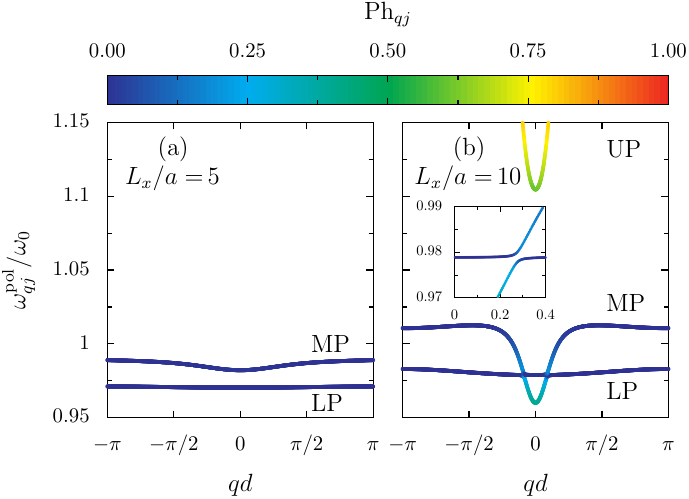}
 \caption{Polaritonic dispersion relation $\omega^\mathrm{pol}_{qj}$ in units of the bare emitter frequency $\omega_0$ in the first Brillouin zone $qd \in [-\pi,+\pi]$.
 The cavity height $L_x$, which tunes the photonic mode frequencies and the light-matter coupling [see Eqs.~\eqref{eq:photonic dispersion} and \eqref{eq:light-matter coupling functions A and B}], is fixed to (a) $L_x/a=5$ and (b) $L_x/a=10$.
 The color code indicates the photonic weight  $\mathrm{Ph}_{qj}$ [given in Eq.~\eqref{eq:app:photonic part fourier}] of each eigenstate, from $0$ (fully dipolar dark state) to $1$ (fully photonic state).
 Intermediate values, visible in panel (b), correspond to polaritonic, hybrid light-matter eigenstates.
 In the figure, the dimerization $|\epsilon|=0.25$, and, as in the remaining of this paper, the lattice constant $d/a=8$, the dimensionless dipole strength $\omega_0 a/c=0.1$, and the aspect ratio of the cavity is fixed to $L_y/L_x=3$.
 }
 \label{fig:Dispersion Polariton}
\end{figure}

We present the polaritonic band structure $\omega^\mathrm{pol}_{qj}$ in units of the bare emitter frequency $\omega_0$ for two different cavity heights $L_x/a$ in Fig.~\ref{fig:Dispersion Polariton}, with a color code indicating the photonic weight $\mathrm{Ph}_{qj}$ of each eigenstate [see Eq.~\eqref{eq:app:photonic part fourier}].
Dark blue states indicate dipolar excitations almost uncoupled to cavity photons, i.e., dark states, light bluish and greenish states indicate mixed, hybrid light-matter polaritonic excitations, while the yellow and orange states are the mainly photonic excitations.
We consider a chain of emitters with a dimerization $|\epsilon|=0.25$, so that the dipolar Hamiltonian \eqref{eq:H_dp with images} has a two-band gapped eigenspectrum.
As in the remaining of the paper, the lattice constant in units of the dipole length scale is set to $d/a=8$.

In Fig.~\ref{fig:Dispersion Polariton}(a), a cavity height $L_x/a=5$ leads the photonic frequencies \eqref{eq:photonic dispersion} to be highly off-resonant with the dipolar ones, so that the two dipolar bands are only slightly renormalized by the cavity photons, and no mixing takes place between dipolar and photonic degrees of freedom, all the dipolar states remaining dark.
Only the two lower bands (LP and MP) are visible on the scale of the figure, the UP one being too high in frequency.
Such a small cavity height does not lead to the strong coupling regime and its associated polaritons, but it does however induce a redshift of the two dipolar bands from the bare emitter frequencies $\omega_0$, which originates from image dipoles due to the cavity walls.

We increase the cavity height to $L_x/a=10$ in Fig.~\ref{fig:Dispersion Polariton}(b), so that the photonic dispersion \eqref{eq:photonic dispersion} approaches the dipolar ones.
In that case, one observes the emergence of polaritonic states around $q=0$, indicated by light blue and green colors in the whole of the LP, MP, and UP bands.
An avoided crossing scheme between the UP and MP branches---a typical signature of the strong coupling regime---is also visible, with the redshifting of the two lower LP and MP bands around $q=0$.
We note that a second avoided crossing between the two lower bands [see the inset in Fig.~\ref{fig:Dispersion Polariton}(b)] is also present, as discussed in Refs.~\cite{Downing2019,Allard2023}.

The dispersion relations displayed in Fig.~\ref{fig:Dispersion Polariton} are for a given value of the (absolute value of the) dimerization parameter $|\epsilon|$. Increasing (decreasing) $|\epsilon|$ results in a widening (narrowing) of the energy gap between the LP and MP bands visible in Fig.~\ref{fig:Dispersion Polariton}(a). Consequently, a larger (smaller) value of $L_x/a$ is required to observe the band modification visible in Fig.~\ref{fig:Dispersion Polariton}(b).

\subsection{Effective dipolar Hamiltonian}

To further simplify our discussion of the Hamiltonian \eqref{eq:total Hamiltonian}, we next perform a Schrieffer-Wolff unitary transformation, a procedure which we detail in Appendix~\ref{sec:Appendix Schrieffer-Wolff}, and that allows us to perturbatively integrate the photonic degrees of freedom of the cavity.
This perturbative scheme provides us with an effective dipolar Hamiltonian containing the indirect effects of the cavity photons on the emitter bare frequencies as well as on the dipolar couplings.
Such an effective Hamiltonian reads
\begin{align}
	\tilde{H}^\mathrm{im}_\mathrm{dp} =&\; \sum_{m=1}^\mathcal{N} 
    \left( \hbar \tilde{\omega}_m^{\mathrm{im},A}  a_{m}^{\dagger} a_{m}^{\phantom{\dagger}} + \hbar \tilde{\omega}_m^{\mathrm{im},B} b_{m}^{\dagger} b_{m}^{\phantom{\dagger}}\right)
    \nonumber \\ 
    &+ \frac{\hbar\Omega}{2}\sum_{\substack{m,m'=1\\(m\neq m')}}^{\mathcal{N}}  \left[ \tilde{f}_{m,m'}^{\mathrm{im},A} \left( a_{m^{\phantom{}}}^{\dagger} a_{m^\prime}^{\phantom{\dagger}} + \mathrm{H.c.} \right) \right. \nonumber \\
    &\left. + \; \tilde{f}_{m,m'}^{\mathrm{im},B} \left( b_{m^{\phantom{}}}^{\dagger} b_{m^\prime}^{\phantom{\dagger}} + \mathrm{H.c.} \right) \right]
    \nonumber \\
    &+ \hbar\Omega\sum_{m,m'=1}^{\mathcal{N}} \tilde{g}^\mathrm{im}_{m,m'} \left( a_{m^{\phantom{}}}^{\dagger} b_{m^\prime}^{\phantom{\dagger}} + \mathrm{H.c.} \right),
\label{eq:SW H_dp with images}
\end{align}
where the onsite frequencies $\omega_m^{\mathrm{im},A}$ and $\omega_m^{\mathrm{im},B}$, the intrasublattice sums $f_{m,m'}^{\mathrm{im},A}$ and $f_{m,m'}^{\mathrm{im},B}$, as well as the intersublattice sum $g_{m,m'}^\mathrm{im}$ of the dipolar Hamiltonian \eqref{eq:H_dp with images}, already renormalized by image dipoles, are further renormalized by the cavity photons as
\begin{subequations}
\label{eq:renormalization by the cavity photons}
    \begin{equation}
    \label{eq:cuboid SW renormalized omega0A}
        \tilde{\omega}_m^{\mathrm{im},A} = \omega_m^{\mathrm{im},A} - \sum_{n_z=1}^{N_z} \frac{\left( \xi^{A}_{mn_z} \right)^2}{ \omega_{n_z}^\mathrm{ph} - \omega_m^{\mathrm{im},A} },
    \end{equation}
    \begin{equation}
        \tilde{\omega}_m^{\mathrm{im},B} = \omega_m^{\mathrm{im},B} - \sum_{n_z=1}^{N_z} \frac{\left( \xi^{B}_{mn_z} \right)^2}{ \omega_{n_z}^\mathrm{ph} - \omega_m^{\mathrm{im},B} },
    \label{eq:cuboid SW renormalized omega0B}
    \end{equation}
    \begin{equation}
        \tilde{f}_{m,m'}^{\mathrm{im},A} = f_{m,m'}^{\mathrm{im},A} - \frac{1}{\Omega}\sum_{n_z=1}^{N_z} \frac{\xi^{A}_{mn_z} \xi^{A}_{m'n_z}}{ \omega_{n_z}^\mathrm{ph} - \omega_m^{\mathrm{im},A} },
    \label{eq:cuboid SW renormalized intrasublatticeA}
    \end{equation}
    \begin{equation}
        \tilde{f}_{m,m'}^{\mathrm{im},B} =f_{m,m'}^{\mathrm{im},B} - \frac{1}{\Omega}\sum_{n_z=1}^{N_z} \frac{\xi^{B}_{mn_z} \xi^{B}_{m'n_z}}{ \omega_{n_z}^\mathrm{ph} - \omega_m^{\mathrm{im},B} },
    \label{eq:cuboid SW renormalized intrasublatticeB}
    \end{equation}
and
    \begin{equation}
        \tilde{g}_{m,m'}^\mathrm{im} = g^{\mathrm{im}}_{m,m'} - \frac{1}{2\Omega}\sum_{n_z=1}^{N_z} \left( \frac{\xi^{A}_{mn_z} \xi^{B}_{m'n_z}}{ \omega_{n_z}^\mathrm{ph} - \omega_m^{\mathrm{im},A} } + \frac{\xi^{A}_{m n_z} \xi^{B}_{m' n_z}}{ \omega_{n_z}^\mathrm{ph} - \omega_{m'}^{\mathrm{im},B} } \right).
    \label{eq:cuboid SW renormalized intersublattice}
    \end{equation}
\end{subequations}
We verified that such a perturbation theory is valid as long as the photonic and dipolar subspaces are well separated, i.e., for $\omega^\mathrm{ph}_{n_z} > \omega^{ \mathrm{im},A/B}_m$.
This limits the values of cavity heights to $L_x/a \lesssim 10$.

As visible in the above expressions, the light-matter coupling induces renormalizations which are increasing with the light-matter coupling functions \eqref{eq:light-matter coupling functions A and B}, and decreasing with the detuning between the cavity photons and the dipolar excitations $\omega^\mathrm{ph}_{n_z} - \omega^{\mathrm{im},A/B}_m$.
Moreover, the cavity modifies nonuniformly the emitter frequencies as well as the dipolar couplings, because of the lack of translation invariance induced by the cavity walls in the $z$ direction.
In the following Sec.~\ref{sec:transverse}, we will study in detail this effect and its implication for the system properties.

\section{Effects of the transverse cavity mirrors}
\label{sec:transverse}

Now that our model as been presented and discussed in the preceding section, let us start our study of the effects induced by the cavity mirrors by investigating the specific role of the transverse cavity walls, namely, the boundary conditions imposed by the cavity in the direction of the chain of emitters.
To this end, we make use of the effective Hamiltonian \eqref{eq:SW H_dp with images} and vary the distance 
$d_\mathrm{cav}$ between the first and last emitters of the chain and the cavity mirrors (cf.\ Fig.~\ref{fig:Sketch}).
For large chains ($\mathcal{N} \gg 1$), such a change in 
the cavity length results only in a marginal modification of the cavity volume, and thus of the photonic degrees of freedom.
Nevertheless, we shall see in the remaining of this section that modifying $d_\mathrm{cav}$ leads to substantial changes in the system properties, solely induced by the cavity mirrors.

\subsection{Mirrors act as defects: Emergence of Tamm states in a regular chain}
\label{sec:Tamm}

We start by considering a regular chain with no dimerization ($d_1=d_2$, $\epsilon=0$).
In such a regular chain of dipolar emitters coupled to a cuboidal multimode cavity, Downing and Mart\'in-Moreno \cite{Downing2021} recently unveiled the appearance of Tamm states, nontopological exponentially localized edge states which are known to originate from defects at the system edges \cite{Tamm1932,Ohno1990}.
To explain the origin of such ``polaritonic Tamm states,'' as coined by the authors of Ref.~\cite{Downing2021}, 
two mechanisms were proposed: (i) the mixing between dipolar and photonic excitations into polaritons and (ii) the possible role of the dipolar coupling beyond nearest neighbors.

Here, using the effective Hamiltonian presented in Sec.~\ref{sec:Model}, we demonstrate that, in fact, the origin of these Tamm states stems from the boundary conditions imposed by the cavity walls in the $z$ direction 
on the electromagnetic field inside the cavity. We show that such edge states exist only in the case of cavity walls in the close vicinity of the ends of the chain of emitters.
Indeed, as we argue below, transverse cavity mirrors close enough to the first and last sites of the chain may induce edge defects on the onsite energies of the emitters as well as in the couplings between them.
We point out that although we consider in this section a regularly spaced chain of emitters with $d_1=d_2$, very similar Tamm states are also found for a dimerized chain, in particular in the trivial phase of the SSH model, when edge states with a topological origin are absent.

\subsubsection{Mirror-induced edge defects}

\begin{figure}[t]
 \includegraphics[width=\linewidth]{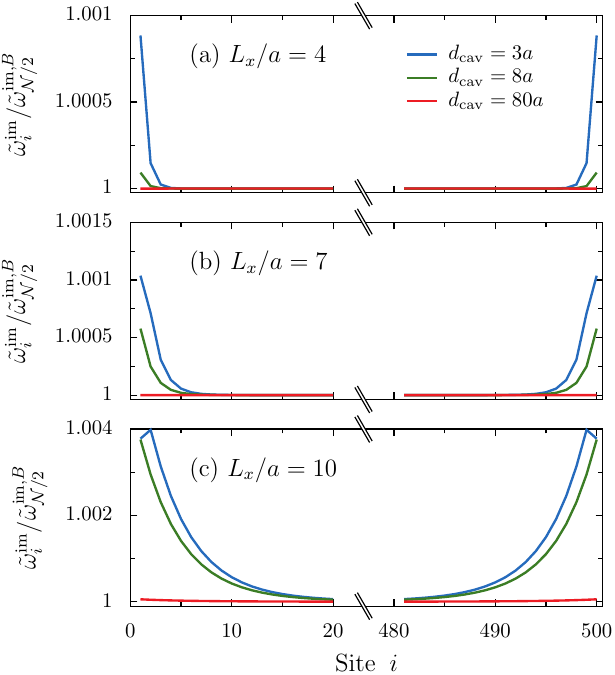}
 \caption{Renormalized onsite frequency $\tilde{\omega}^\mathrm{im}_i$ [see Eqs.~\eqref{eq:cuboid SW renormalized omega0A} and \eqref{eq:cuboid SW renormalized omega0B}] scaled by its value close to the middle of the chain  $\tilde{\omega}^{\mathrm{im},B}_{\mathcal{N}/2}$, and as a function of the sites $i$ along the chain.
 Increasing values of the distance to the transverse cavity walls $d_\mathrm{cav}/a$ are considered from blue to green and red lines.
 The height of the cavity is (a) $L_x/a = 4$, (b) $L_x/a = 7$, and (c) $L_x/a = 10$.
 The chain under consideration in the ﬁgure is composed of $2\mathcal{N} = 500$ emitters and has no dimerization ($\epsilon=0$).}
 \label{fig:Renormalized onsite frequency}
\end{figure}

The above-mentioned effect can be understood by studying the influence of the distance to the transverse cavity walls, $d_\mathrm{cav}$, on the renormalized onsite frequencies \eqref{eq:cuboid SW renormalized omega0A} and \eqref{eq:cuboid SW renormalized omega0B}.
Importantly, due to the broken translation invariance induced by both the image dipoles and the light-matter coupling, the latter frequencies are site dependent.
We represent the renormalized onsite frequencies $\tilde{\omega}^{\mathrm{im}, A}_m$ and $\tilde{\omega}^{\mathrm{im}, B}_m$ along the sites $i$ of a chain of $500$ emitters in Fig.~\ref{fig:Renormalized onsite frequency}.
In order to facilitate the following discussion, we recast the cell-dependent quantities \eqref{eq:cuboid SW renormalized omega0A} and \eqref{eq:cuboid SW renormalized omega0B} in a site-dependent renormalized frequency $\tilde{\omega}^\mathrm{im}_i$. In order to highlight the dependency of $\tilde{\omega}^\mathrm{im}_i$ on $i$, we display this quantity scaled by its value close to the middle of the chain $\tilde{\omega}^{\mathrm{im},B}_{\mathcal{N}/2}$. 
The blue, green, and red lines in Fig.~\ref{fig:Renormalized onsite frequency} represent increasing values of $d_\mathrm{cav}/a=3$, $8$, and $80$, respectively.

In the case of a small cavity height $L_x/a=4$ and a small distance $d_\mathrm{cav}/a=3$ [see the blue line in Fig.~\ref{fig:Renormalized onsite frequency}(a)], we observe extremely sharp peaks of the onsite frequencies at the first and last sites of the chain, exactly as in the presence of defects.
The bare dipole frequencies being globally redshifted by the presence of the cavity, this means that dipoles at the edges of the chain are less impacted by such a redshift.
This lack of redshift can be understood from the expression of the light-matter coupling functions \eqref{eq:light-matter coupling functions A and B}.
Because of the hard wall boundary conditions in the $z$ direction, the electromagnetic field which couples to the emitters must vanish on the cavity mirrors, so that Eq.~\eqref{eq:light-matter coupling functions A and B} corresponds to a sine function of the product of the wave number $k_z$ and the emitter position $z_m^{A/B}$.
This leads to the fact that, for a small value of the position $z_m^{A/B}$ of the emitters, the coupling to photons with a small wave number $k_z$, namely, with the ones which are the less detuned with dipoles, approaches $0$.
Such a situation arises only in the case of a small distance $d_\mathrm{cav}$, for which the first and last emitters are close to the points of vanishing electromagnetic field. 
As can be seen in Fig.~\ref{fig:Renormalized onsite frequency}(a) (green and red lines),
the defects at the edges of the chain indeed fade out for larger values of $d_\mathrm{cav}$.
We note that in Ref.~\cite{Downing2021}, a value of $d_\mathrm{cav}=3a$ was considered.

Increasing the cavity height in Figs.~\ref{fig:Renormalized onsite frequency}(b) and \ref{fig:Renormalized onsite frequency}(c)---and hence approaching resonance between dipolar and photonic excitations---produces an expansion of the defect on many sites near the two edges.
Such smoothening of the defect on a large number of sites originates from the increasing renormalization due to the cavity [see Eqs.~\eqref{eq:cuboid SW renormalized omega0A} and \eqref{eq:cuboid SW renormalized omega0B}], so that the lack of redshift at the edges we discussed above is already visible farther from the cavity walls.
Moreover, the larger effect of the cavity near resonance is also reflected through the significant increase of the magnitude of the defect from Fig.~\ref{fig:Renormalized onsite frequency}(b) to Fig.~\ref{fig:Renormalized onsite frequency}(c).

One also notes in Fig.~\ref{fig:Renormalized onsite frequency}(c) (blue line) a decrease of the defect on the very first and very last sites, in the case of a very small value of $d_\mathrm{cav}$.
This reflects the combined effect of the light-matter coupling and of image dipoles, both inducing inhomogeneities in the effective onsite frequencies.
While the renormalizations arising from the light-matter coupling present the dominant contributions, in the case of very close transverse mirrors image dipoles may compete through an opposite effect.
Finally, while it is not visible on Fig.~\ref{fig:Renormalized onsite frequency} due to the considered normalization of the frequencies $\tilde{\omega}_i^\mathrm{im}$, we also notice a global redshift of the onsite frequencies as the distance $d_\mathrm{cav}$ or the number of emitters increases.

\subsubsection{Tamm states}

To assess the effects of the transverse cavity mirrors on the emergence of Tamm states, we next compute the participation ratio (PR) of the eigenstates of the system.
Such a quantity, defined as
\begin{equation}
    \mathrm{PR}(n) = \frac{ 1  }{ \sum_{i=1}^{2\mathcal{N}}|\Psi_i(n)|^4   },
\label{eq:Participation Ratio}
\end{equation}
with $\Psi_i(n)$ the eigenstates of the Hamiltonian \eqref{eq:SW H_dp with images}, provides information about the typical number of sites $i$ occupied by an eigenstate $n$.
In the one-dimensional system considered in this work, it is thus proportional to the localization length of an eigenstate.

\begin{figure}[t]
 \includegraphics[width=\linewidth]{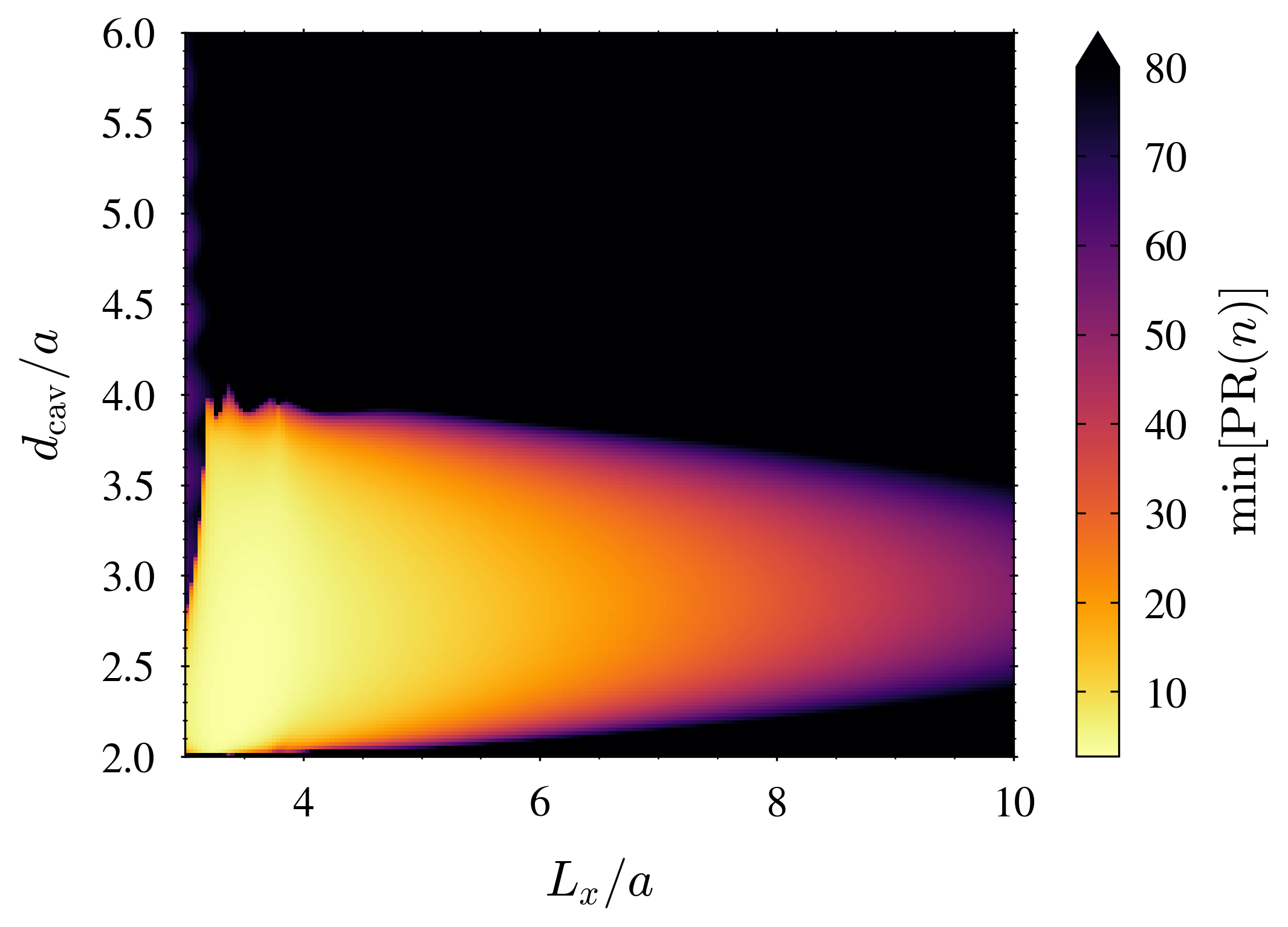}
 \caption{Participation ratio of the most localized state as a function of both the distance to the transverse cavity walls $d_\mathrm{cav}/a$ and the cavity height $L_x/a$, for a chain of $2\mathcal{N}=500$ emitters.
 A typical delocalized state has a PR $= 4\mathcal{N}/3 \simeq 333$.
 The considered chain has no dimerization ($\epsilon=0$).}
 \label{fig:Heatmap dcav - Lx}
\end{figure}

We present in Fig.~\ref{fig:Heatmap dcav - Lx} the minimum value of the PR found in the system as a function of both the cavity height $L_x/a$ and the distance to the transverse walls $d_\mathrm{cav}/a$.
A chain of $2\mathcal{N}=500$ emitters is considered, for which a typical delocalized state has a PR $\simeq 4\mathcal{N}/3 \simeq 333$.
In this way, the presence of an eigenstate with a small value of PR $\lesssim 60$, revealed in Fig.~\ref{fig:Heatmap dcav - Lx} through a bright color, indicates that the chain hosts a localized state.

By examining the density plot of Fig.~\ref{fig:Heatmap dcav - Lx}, one observes that the parameter regime in which the system hosts localized states is in accordance with the above discussion on the mirror-induced defects at the edges of the chain.
Indeed, Tamm states are visible only for small values of $d_\mathrm{cav}/a\lesssim 4$ (i.e., for values of the order of the interemitter distance), and are localized on a growing number of sites as the cavity height $L_x/a$ increases, in a similar manner as the defects we discussed in Fig.~\ref{fig:Renormalized onsite frequency}.

\begin{figure}[t]
 \includegraphics[width=\linewidth]{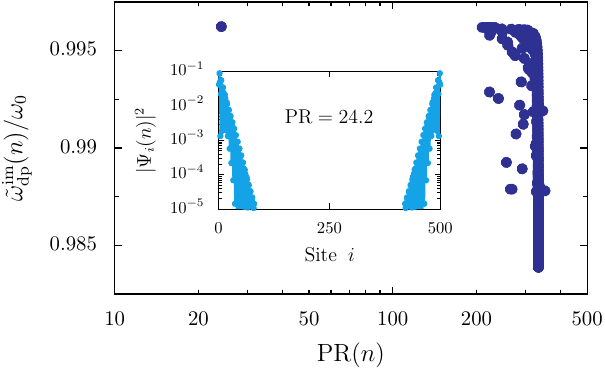}
 \caption{Participation ratio $\mathrm{PR(n)}$ and eigenfrequency $\tilde{\omega}_\mathrm{dp}^\mathrm{im}(n)/\omega_0$ of each eigenstate $n$ of the system.
 Two degenerate eigenstates with a very small PR and a frequency slightly above the rest of the spectrum indicate the presence of Tamm states.
 The inset shows the probability density along the chain for one of these two Tamm states.
 A chain of $2\mathcal{N}=500$ dipoles with no dimerization ($\epsilon=0$) and embedded in a cavity with height $L_x/a=7$ and distance $d_\mathrm{cav}/a=3$ is considered.}
 \label{fig:Freq vs PR and Proba density}
\end{figure}

We display in Fig.~\ref{fig:Freq vs PR and Proba density} the eigenspectrum $\tilde{\omega}_\mathrm{dp}^\mathrm{im}(n)/\omega_0$ of the Hamiltonian \eqref{eq:SW H_dp with images} along with the corresponding PR of each eigenstate $n$, in the case of a system hosting Tamm edge states.
For this purpose, we set the cavity height to $L_x/a=7$ and the distance to the transverse cavity walls to $d_\mathrm{cav}/a=3$.
Two degenerate eigenstates located on top of the eigenspectrum, i.e., within the polaritonic gap, show a very small value of $\mathrm{PR} \simeq 24.2$, in contrast to all the other eigenstates of the system.
The inset of Fig.~\ref{fig:Freq vs PR and Proba density} shows the probability density $|\Psi_i(n)|^2$ of one of these two degenerate states as a function of the site $i$ along the chain.
A clear exponential localization at the two edges of the chain ensures the fact that such an eigenstate is indeed a localized Tamm state.

To conclude this section on the emergence of Tamm states, we note that the mirror-induced defects on the onsite energies discussed in Fig.~\ref{fig:Renormalized onsite frequency} cannot solely explain the entire results displayed in Fig.~\ref{fig:Heatmap dcav - Lx}.
Indeed, as already mentioned above, very similar defects at the edges of the chain are also present in the coupling constants \eqref{eq:cuboid SW renormalized intrasublatticeA} to \eqref{eq:cuboid SW renormalized intersublattice}, and are essential to the formation of Tamm states.
We also note that Tamm states can be found neglecting image dipoles \cite{Downing2021}. Such an approximation significantly modifies the results of  Fig.~\ref{fig:Heatmap dcav - Lx} (not shown), notably by increasing the range of $d_\mathrm{cav}$ for which Tamm states are present.

\subsection{Mirrors assist localization: Protection of edge states in a dimerized chain}
\label{sec:Edge states}

We now discuss the interplay of topology and strong light-matter coupling in our system. To this end, we consider a dimerized chain of dipolar emitters, with the parameter $\epsilon \neq 0$.
Such dimerization, leading to alternating nearest neighbor couplings between the dipoles, makes the system analogous to the SSH model of one-dimensional topological physics \cite{Su1979,Asboth2016}.
In the case of a dimerization $\epsilon>0$, so that $d_1 > d_2$, the quasistatic dipolar chain in the absence of a cavity hosts edge states, similarly to the ones of the original SSH model of polyacetylene \cite{Downing2017_Topological,Pocock2018,Downing2018}.\footnote{We note, however, that from the broken chiral symmetry induced by the all-to-all quasistatic dipole couplings \eqref{eq:app:dipolar couplings}, the edge states hosted in such a dipolar SSH model are not symmetry-protected topological edge states.}

The coupling to the cavity, by effectively renormalizing the dipole-dipole couplings [see Eqs.~\eqref{eq:renormalization by the cavity photons}], then turns the system into a polaritonic SSH model, and, depending on the cavity dimensions, the topological properties of the chain may be drastically affected \cite{Downing2019, Allard2023}.
This permits the investigation of the interplay between topological properties and (strong) light-matter coupling, a topic that has recently received considerable attention \cite{Mann2018,Nie2020,Mann2022,PerezGonzalez2022,Wei2022,Appugliese2022,Pirmoradian2023,Zhao2024_arXiv,Enkner2024}.

\begin{figure*}[t]
    \includegraphics[width=\linewidth]{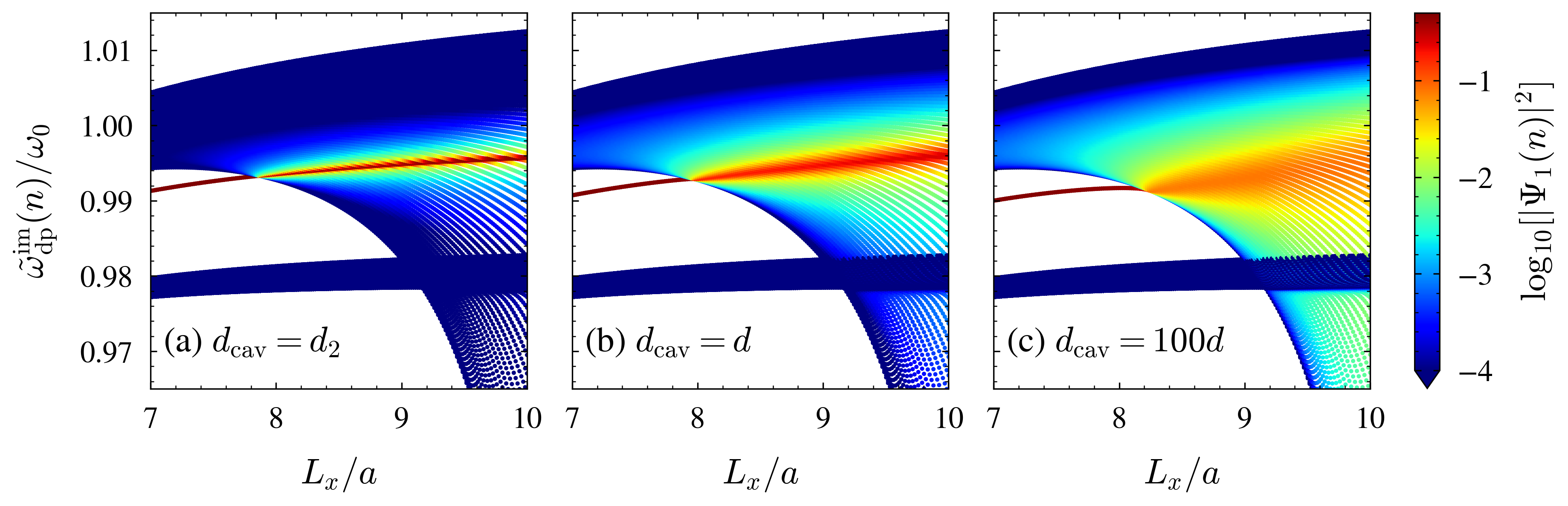}
    \caption{Eigenspectrum  $\tilde{\omega}_\mathrm{dp}^\mathrm{im}(n)/\omega_0$ as a function of the cavity height $L_x/a$ for a chain with a dimerization parameter $\epsilon = 0.25$.
    The color associated to each eigenstate $n$ indicates the logarithm of its probability density on the ﬁrst site $|\Psi_1(n)|^2$.
    A reddish (bluish) color denotes a state highly (poorly) localized on the edges of the chain.
    Panels (a)--(c) show increasing distances to the transverse cavity walls $d_\mathrm{cav}$, leading to the diffusion of polaritonic edge states into the bulk.
    The number of dimers $\mathcal{N}=250$.}
    \label{fig:Freq vs Lx for different dcav}
\end{figure*}

The polaritonic SSH model \eqref{eq:total Hamiltonian}, sketched in Fig.~\ref{fig:Sketch}, has been first studied in Ref.~\cite{Downing2019}, in the case of a cuboidal cavity with transverse cavity walls at a distance $d_\mathrm{cav} = d - d_1/2$, very close to the chain of dipoles.
In the topological phase of the original SSH model ($\epsilon>0$), Downing \textit{et al}.\ unveiled a transition from in-gap dipolar edge states to hybrid polaritonic edge states which profit from both exponential edge localization and bulk delocalization.
This transition arises from the filling of the energy gap by polaritonic states as the cavity approaches resonance with the dipolar emitters (see the band structure in Fig.~\ref{fig:Dispersion Polariton}), leading edge states to mix with bulk polaritons.

A further study of the Hamiltonian \eqref{eq:total Hamiltonian} has been realized by the authors in Ref.~\cite{Allard2023}, in which a waveguide cavity, with $d_\mathrm{cav} \to \infty$, has been assumed.
In such a geometry, the mixing of edge states with bulk polaritons has been shown to be strongly increased, with the emergence of multiple polaritonic edge states occupying a large portion of the frequency spectrum.
Detailed energy transport simulations have been carried out in Ref.~\cite{Allard2023}, demonstrating efficient edge-to-edge propagation through such highly hybridized polaritonic edge states.

Recently, an experimental realization of the polaritonic SSH model \eqref{eq:total Hamiltonian} has been achieved in Ref.~\cite{Zhao2024_arXiv}, using macroscopic microwave helical resonators embedded in a cuboidal copper cavity.
The configuration proposed in Ref.~\cite{Downing2019} was chosen in the experiment, with cavity walls close to the ends of the emitter chain, and transport measurements confirmed edge-to-edge propagation.

In this section, we analyze the precise influence of the distance $d_\mathrm{cav}$ to the cavity transverse walls on the formation and properties of polaritonic edge states, and explain its impact by means of the effective Hamiltonian \eqref{eq:SW H_dp with images}.
To this end, we numerically diagonalize the latter Hamiltonian considering a dimerization $\epsilon=0.25$ and several values of the distance $d_\mathrm{cav}$.
For each value, we compute the eigenfrequencies $\tilde{\omega}_\mathrm{dp}^\mathrm{im}(n)$ as a function of the cavity height $L_x/a$ in order to study the effect of the cavity as it approaches resonance with the dipolar emitters.
The results are presented in Fig.~\ref{fig:Freq vs Lx for different dcav}, in which we also indicate as a color code the logarithm of the probability density $|\Psi_1(n)|^2$ of the eigenstate $n$ on the first site $i=1$ of the chain, from blue (delocalized) to red (localized on the edges).

In Fig.~\ref{fig:Freq vs Lx for different dcav}(a), a small distance $d_\mathrm{cav}=d_2=3a$ is assumed, similarly to the case considered in Ref.~\cite{Downing2019}.
On the left of the panel, for a cavity height $L_x/a=7$, the photons are off-resonant with the dipolar emitters and two nearly degenerate dipolar edge states are visible in red within the band gap.
Increasing the cavity height up to $L_x/a=10$ and hence approaching resonance between photonic and matter degrees of freedom, i.e., approaching the strong light-matter coupling, the bottom of the upper band is redshifted and the band gap is filled with polaritons, as we discussed for the band structure of Fig.~\ref{fig:Dispersion Polariton}.
Once such a redshift attains the dipolar edge states, an hybridization between the latter and bulk states occurs, leading polaritons to inherit edge localization.
Interestingly, with transverse walls that close from the ends of the chain, we observe a strong persistence of the edge state in the strong-coupling regime, with few polaritons being impacted by the edge states, and the latter conserving their very high edge localization with $|\Psi_1(n)|^2 \sim 0.5$. 

Increasing the distance to the transverse cavity walls to $d_\mathrm{cav}=d$ and $d_\mathrm{cav}=100d$ in Figs.~\ref{fig:Freq vs Lx for different dcav}(b) and \ref{fig:Freq vs Lx for different dcav}(c), respectively, we observe a continuous enhancement of the diffusion of edge states into bulk polaritons in the strong-coupling regime, through the spreading of a red spot in the right of the plots.
Moving away the transverse cavity walls therefore leads to the formation of an increasing number of polaritonic edge states, which occupy a growing portion of the eigenspectrum, and which are increasingly less localized on the edges.

\begin{figure}[t]
 \includegraphics[width=\linewidth]{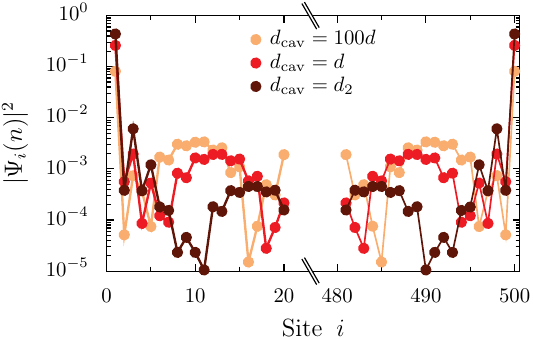}
 \caption{Probability density $|\Psi_i(n)|^2$ along the sites $i$ of a chain composed of $\mathcal{N} = 250$ dimers, for the polaritonic edge state $n$ with the lowest participation ratio.
 Increasing distances $d_\mathrm{cav}$ are considered, demonstrating the enhanced localization of edge states for dipolar chains close to the transverse cavity walls.
 The cavity height $L_x/a = 10$ and the dimerization $\epsilon = 0.25$.}
 \label{fig:Edge states for different dcav}
\end{figure}

To gain some insight about the effect of the distance $d_\mathrm{cav}$ on polaritonic edge states, we next consider the same chain as in Fig.~\ref{fig:Freq vs Lx for different dcav} and set the cavity height to $L_x/a=10$, so that the system is in the strong-coupling regime and host polaritonic edge states.
We display in Fig.~\ref{fig:Edge states for different dcav} the probability density along the chain of the most localized of these polaritonic edge states for increasing values of the distance $d_\mathrm{cav}$, corresponding to the ones of Fig.~\ref{fig:Freq vs Lx for different dcav}.
In all three cases, the polaritonic edge states are formed by an exponential localization on the edges of the chain and a delocalized oscillating component in the bulk.
Importantly, however, the exponential localization is diminished as the distance to the transverse cavity walls $d_\mathrm{cav}$ increases, while at the same time the bulk delocalization intensifies.
Indeed, the probability density on the first and last sites is divided by a factor of $5$ as we move away the transverse walls from $d_\mathrm{cav}=d_2$ to $d_\mathrm{cav}=100d$, whereas the amplitude in the bulk of the chain increases by one order of magnitude.
In that way, nearby transverse cavity walls protect the edge state localization, but decrease the bulk delocalization which allows efficient edge-to-edge transport \cite{Allard2023}.
The distance $d_\mathrm{cav}$ thus may serve as a knob to tune the properties of topological edge states in the strong light-matter coupling regime.

Two different, mutually reinforcing effects can be identified to explain this phenomenon.
First, we have already seen in Sec.~\ref{sec:Tamm} that in the case of a small distance $d_\mathrm{cav}$, dipoles located around the ends of the chain are less affected by the light-matter coupling than those in the bulk of the chain due to the vanishing of the electromagnetic field on the cavity mirrors.
This, therefore, protects the edge localization.
Second, pushing away the transverse mirrors also enhances the long-range coupling between the dipolar emitters, thus increasing the coupling between the edges and the bulk of the chain.

\begin{figure}[t]
 \includegraphics[width=\linewidth]{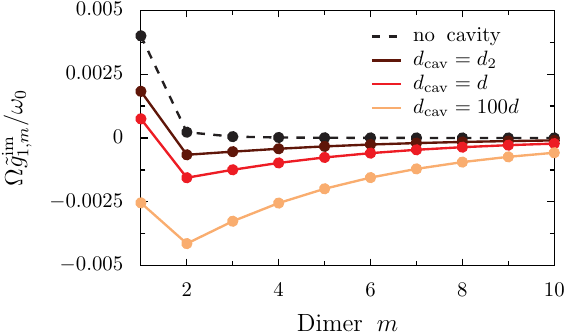}
 \caption{Intersublattice coupling strength $\Omega \tilde{g}^\mathrm{im}_{1,m}/\omega_0$ [see Eq.~\eqref{eq:cuboid SW renormalized intersublattice}], i.e., coupling between the first dipole of the chain (located on a site $A$) and the $m$th dipole located on a site $B$.
 The solid colored lines show increasing values of the distance to the transverse cavity walls $d_\mathrm{cav}$, while the black dashed line indicates the case of a quasistatic dipolar chain without cavity [see Eq.~\eqref{eq:app:dipolar coupling g}].
 The parameters are the same as in Fig.~\ref{fig:Edge states for different dcav}.}
 \label{fig:g_1_m for different dcav}
\end{figure}

We exemplify the latter effect in Fig.~\ref{fig:g_1_m for different dcav} by showing the strength of the renormalized intersublattice coupling in units of the bare dipole frequency $\Omega\tilde{g}^\mathrm{im}_{1,m}/\omega_0$ [see Eq.~\eqref{eq:cuboid SW renormalized intersublattice}] as a function of the dimer index $m$.
Such a quantity represents the renormalized dipole-dipole coupling between the first dipole of the chain (located on the sublattice $A$) and the $m$th dipole located on a site $B$.
As in Fig.~\ref{fig:Edge states for different dcav}, we set the cavity height to $L_x/a=10$ and we look at the same increasing values of the distance $d_\mathrm{cav}$, from $d_\mathrm{cav}=d_2$ (brown line), to $d_\mathrm{cav}=d$ (red line), and $d_\mathrm{cav}=100d$ (beige line).
We also display in Fig.~\ref{fig:g_1_m for different dcav} the usual quasistatic dipole-dipole coupling in the absence of a cavity [see Eq.~\eqref{eq:app:dipolar coupling g}] as a black dashed line.
The quasistatic coupling is increasingly renormalized as one moves away the transverse mirrors from the chain, inducing a drastic enhancement of the coupling between the first emitter and its neighbors. 
Such an increased long-distance coupling reduces the edge localization, and enlarges the bulk delocalization, as is visible in Fig.~\ref{fig:Edge states for different dcav}.

To conclude, in this section we investigated the influence of the transverse cavity walls on the edge states of a dimerized chain of dipoles in a cavity.
While it is known that in the strong light-matter coupling regime edge states couple to polaritons, leading to hybrid polaritonic edge states \cite{Downing2019,Allard2023}, here we unveiled that the properties of such polaritonic states are drastically affected by the distance $d_\mathrm{cav}$ between the first and last dipoles and the transverse cavity walls.
Recent experiments have demonstrated the possibility of edge-to-edge energy transport using such hybrid states in the case of nearby transverse cavity mirrors \cite{Zhao2024_arXiv}. Our results suggest that moving away the mirrors from the chain should allow for a more efficient edge-to-edge transport.

\section{Effect of the longitudinal cavity mirrors: image-induced topological phase transition}
\label{sec:Images}

In the previous Sec.~\ref{sec:transverse}, we explored the particular importance of the transverse cavity walls in the formation of different types of edge states at the ends of the system.
We now examine the specific role of the longitudinal cavity mirrors.
To this end, in this section we assume a large enough distance $d_\mathrm{cav} = 100d$, so that the effect of transverse cavity mirrors can be disregarded.

We have seen in Sec.~\ref{sec:Model} that tuning the spacing between these longitudinal mirrors, through the height $L_x$ and width $L_y$ of the cavity, allows for the resonance between dipolar emitters and photons, and for the strong light-matter coupling regime.
Independently of such a resonance, however, the latter mirrors, through the boundary conditions they impose on the electromagnetic field inside the cavity, may also modify the eigenspectrum of the system.

Indeed, as the cavity walls are considered as perfect metallic plates, both the electric field and potential must vanish on their surfaces.
To account for this new boundary condition in the Poisson problem under consideration, one must add effective, or ``image'' dipoles to the system \cite{Jackson2007}.
Such additional dipoles, located outside of the cavity (see Fig.~\ref{fig:Sketch}), modify the bare dipolar degrees of freedom.
As visible in the dipolar Hamiltonian \eqref{eq:H_dp with images} and discussed in detail in Appendix \ref{sec:Appendix images}, they quench the quasistatic dipole-dipole coupling and redshift the bare dipole frequencies.

The effect of image dipoles increases with the proximity of the cavity walls to the emitters.
In the regime considered in this work, reducing the cavity height $L_x$ (and thus the cavity width $L_y=3L_x$) leads to photonic modes highly detuned from the dipolar emitters.
In this section, we stay in such an off-resonant regime with $\omega^\mathrm{ph}_{n_z} \gg \omega_0$ by considering small cavity heights $L_x/a \lesssim 6$, and demonstrate how in this case image dipoles may lead to drastic changes in the system properties.
In particular, as in Sec.~\ref{sec:Edge states} we consider a dimerized chain of emitter with $d_1 \neq d_2$, mimicking an SSH-like model.
In such a two-band model, we unveil topological phase transitions solely induced by image dipoles, with the disappearance and appearance of in-gap edge states as the cavity cross section is decreased.
We note that very similar band structure renormalizations have also been predicted in honeycomb dipolar metasurfaces embedded in small cavities, both analytically and using exact finite element methods \cite{Mann2018,Mann2020}.

\subsection{Image-induced band-gap closure}

In the off-resonant regime considered here ($L_x/a \lesssim 6$), the light-matter coupling itself has little impact on the eigenspectrum.
Without taking into account the effect of image dipoles, the dimerized chain of emitters then nearly corresponds to the usual SSH model, hosting edge states if and only if $d_1>d_2$  ($\epsilon>0$).
However, the quenching of the dipole-dipole coupling induced by image dipoles leads to the flattening of the two dipolar bands and to a drastic reduction of the band gap, eventually resulting in a band-gap closure.

To reveal this phenomenon, we study the band structure of the system as we decrease the cavity cross section, hence increasing the effect of image dipoles.
To this end, we consider the thermodynamic limit of infinitely long chain and cavity and derive the Fourier space equivalent of the effective dipolar Hamiltonian \eqref{eq:SW H_dp with images}.
As detailed in Appendix \ref{sec:Appendix Fourier transformation}, one then obtains the two-band eigenspectrum $\tilde{\omega}^\mathrm{im}_{\mathrm{dp},q\tau}$ [see Eq.~\eqref{eq:app:bandstructure SW}], with $\tau = +$ ($-$) denoting the upper (lower) band. 
 
\begin{figure}[t]
 \includegraphics[width=\linewidth]{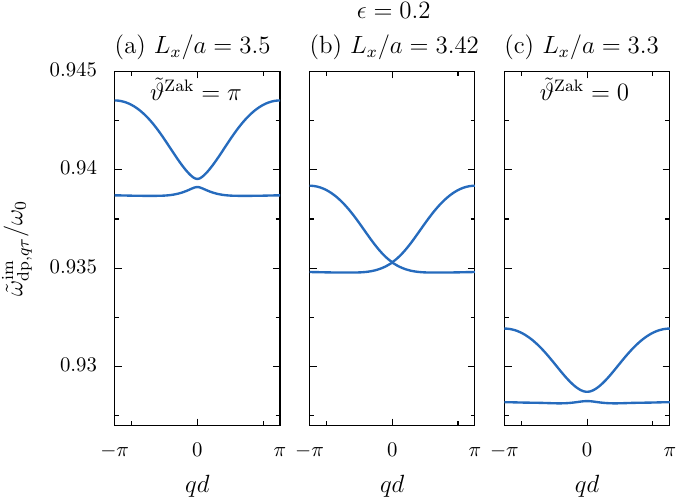}
 \caption{Band structure $\tilde{\omega}^\mathrm{im}_{\mathrm{dp},q\tau}$ [see Eq.~\eqref{eq:app:bandstructure SW}] in units of $\omega_0$ in the first Brillouin zone.
 Three decreasing cavity heights are shown in panels (a)--(c), revealing a band-gap closing at $q=0$ when $L_x/a=3.42$.
 Importantly, except at the transition point, an energy gap is always present between the two bands.
 In the figure, the dimerization $\epsilon=0.2$.}
 \label{fig:Dispersion triple plot TPT with BEC}
\end{figure}
An example of image-induced gap closing is shown in Fig.~\ref{fig:Dispersion triple plot TPT with BEC}, in which the latter band structure is plotted in the first Brillouin zone for decreasing values of the cavity height.
In the figure, the dimerization is fixed to $\epsilon=0.2$.
One observes that going from $L_x/a=3.5$ in Fig.~\ref{fig:Dispersion triple plot TPT with BEC}(a) to $L_x/a=3.3$ in Fig.~\ref{fig:Dispersion triple plot TPT with BEC}(c) leads to the closing and reopening of the band gap at $q=0$, along with a global redshift of the eigenfrequencies.

To ensure that such a band-gap closure is associated with a topological phase transition, we compute a bulk topological invariant of the system, namely, the Zak phase \cite{Delplace2011}
\begin{equation}
    \tilde{\vartheta}^\mathrm{Zak} = \mathrm{i}\int_{-\pi/d}^{+\pi/d}  \mathrm{d}q \braket{\tilde{\psi}^\mathrm{im}_{q\tau} | \partial_q \tilde{\psi}^\mathrm{im}_{q\tau}} \quad \mathrm{mod} \; 2\pi,
\label{eq:Zak phase}
\end{equation}
which we evaluate through the Wilson-loop approach. This method, detailed in Ref.~\cite{Wang_2019}, is based on a discretization of the integral over the momentum $q$, and is particularly suitable for numerical implementation.
Despite the fact that the chiral symmetry, a characteristic symmetry of the original SSH model, is broken by the longer range dipolar coupling induced by the cavity, the system still conserves inversion symmetry allowing the quantity \eqref{eq:Zak phase} to be quantized, and to represents a $\mathbb{Z}_2$ topological invariant of the model \cite{Miert2016}.\footnote{We note that, in contrast to the terminology used in most of the literature, Eq.~\eqref{eq:Zak phase} does not formally represent a Zak phase but rather $\pi$ times a winding number defined with respect to a specific choice of unit cell, as discussed in detail in Ref.~\cite{Fuchs2021}.}
We recall that in the original SSH model, the topological invariant \eqref{eq:Zak phase} equals $\pi$ ($0$), i.e., is nontrivial (trivial), when the dimerization $\epsilon>0$ ($\epsilon<0$).

In Fig.~\ref{fig:Dispersion triple plot TPT with BEC}(a), the Zak phase $\tilde{\vartheta}^\mathrm{Zak} = \pi$, so that the system is in a nontrivial phase, as the original SSH model with the same dimerization $\epsilon=0.2$.
In Fig.~\ref{fig:Dispersion triple plot TPT with BEC}(c), however, $\tilde{\vartheta}^\mathrm{Zak} = 0$, signifying that the image-induced gap closing has led the system to enter into a trivial phase.

As we shall see in the remaining of this section, several band-gap closures and changes of the bulk topological invariant of the system occur as the cavity height is diminished and the effect of image dipoles becomes important.
However, it is important to note that in the system under consideration, namely, a chiral-symmetry broken SSH model, such a change in the Zak phase does not always imply the appearance or disappearance of topological edge states.
Indeed, it has been shown in Ref.~\cite{Downing2019} that the polaritonic SSH model under study does not follow the bulk-edge correspondence (BEC), i.e., the correspondence between a nontrivial bulk topological invariant and the presence of in-gap edge states in the finite system.\footnote{In the terms of the 10-fold way classification, such a one-dimensional system with no chiral nor charge conjugation symmetry belongs formally to a trivial Altland-Zirnbauer class, so that no BEC should be expected \textit{a priori} \cite{Kitaev2009}.}

This breakdown of the BEC that results from the broken chiral symmetry of the model, and that has been predicted in many similar bipartite systems \cite{DiLiberto2014,Longhi2018,PerezGonzalez2019,Pocock2019,Malki2019,Jiao2021}, occurs due to the asymmetry between the two bands.
It leads one of the two bands to absorb the in-gap edge states before closing the gap at a given $q$.

\begin{figure}[t]
 \includegraphics[width=\linewidth]{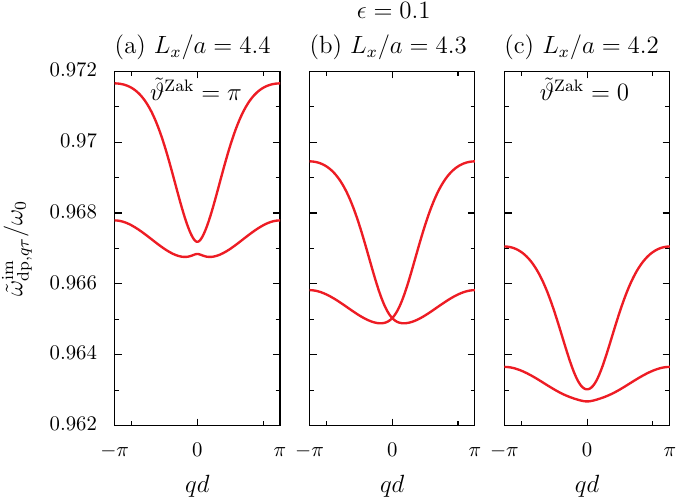}
 \caption{Band structure $\tilde{\omega}^\mathrm{im}_{\mathrm{dp},q\tau}$ [see Eq.~\eqref{eq:app:bandstructure SW}] in units of $\omega_0$ in the first Brillouin zone.
 Three decreasing cavity heights are shown in panels (a)--(c), revealing here a band-gap closing at $q=0$ when $L_x/a=4.3$.
 Importantly, the energy gap is closed in all panels, both before and after the transition point.
 In the figure, the dimerization $\epsilon=0.1$.}
 \label{fig:Dispersion triple plot TPT without BEC}
\end{figure}

We exemplify this phenomenon in Fig.~\ref{fig:Dispersion triple plot TPT without BEC}, in which we plot the band structure again for three decreasing values of the cavity height $L_x/a$, here with a fixed dimerization $\epsilon=0.1$.
In Fig.~\ref{fig:Dispersion triple plot TPT without BEC}(a), in which $L_x/a=4.4$, the Zak phase $\tilde{\vartheta}^\mathrm{Zak}=\pi$ indicates a nontrivial topology, exactly as was the case in Fig.~\ref{fig:Dispersion triple plot TPT with BEC}(a).
An important difference between the two figures, however, is the fact that in Fig.~\ref{fig:Dispersion triple plot TPT without BEC}(a) the energy gap is closed, that is, the system is in a ``metallic'' phase in the language of condensed matter electronic systems.
In such a gapless system, no in-gap edge states can then exist.
Therefore, the band-gap closure at $q=0$ that is visible in Fig.~\ref{fig:Dispersion triple plot TPT without BEC}(b) when $L_x/a=4.3$, while modifying the bulk topological invariant, is not associated with the disappearance of in-gap edge states, since the latter were already absent in Fig.~\ref{fig:Dispersion triple plot TPT without BEC}(a).
We note that such a situation in which the in-gap edge states merge into the upper bulk band was precisely what we discussed in the previous Sec.~\ref{sec:Edge states}.
In the latter case, however, the merging of the edge states into the upper band was induced by the entrance of the system into the strong-coupling regime, so that is was accompanied with polaritons inheriting edge state localization, what we coined polaritonic edge states.

From this lack of reliability of the bulk topology to predict edge physics, it is essential to study a finite system.
We shall see that the image-induced renormalization of the eigenspectrum leads to a complex phase diagram with a variety of topological phase transitions, some respecting the BEC, as the one shown in Fig.~\ref{fig:Dispersion triple plot TPT with BEC}, while others breaking it, as exemplified in Fig.~\ref{fig:Dispersion triple plot TPT without BEC}.

\subsection{Edge states in a finite chain}

\begin{figure}[t]
 \includegraphics[width=\linewidth]{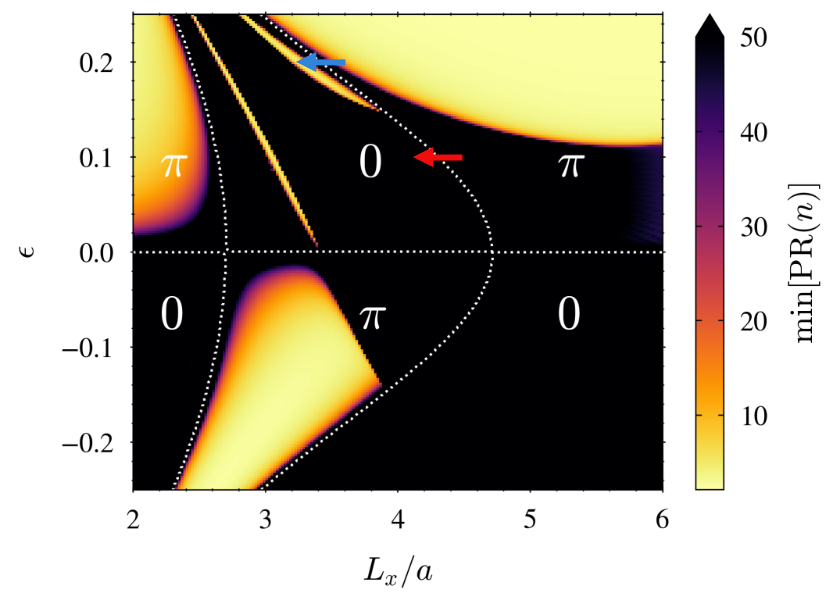}
 \caption{Smallest participation ratio among the eigenstates of the system, as a function of both the dimerization $\epsilon$ and the cavity height $L_x/a$.
 The bright yellow regions do host highly localized edge states, while the black regions do not.
 In the figure, the considered chain is composed of $\mathcal{N}=100$ dimers.}
 \label{fig:Heatmap minPR epsilon vs Lx}
\end{figure}

To obtain a phase diagram of the presence or absence of in-gap edge states in the system, we now diagonalize the Hamiltonian \eqref{eq:SW H_dp with images} of a finite dimerized chain of $\mathcal{N}=100$ dimers.
We compute the participation ratio \eqref{eq:Participation Ratio} of each of the eigenstates of the Hamiltonian, and, similarly as what we have done in Sec.~\ref{sec:Tamm}, we extract the minimal participation ratio hosted by the system.
We show the result of this procedure in Fig.~\ref{fig:Heatmap minPR epsilon vs Lx}, as a function of both the cavity height $L_x/a$ and the dimerization parameter $\epsilon$.
Yellow areas reveal regions of the parameter space for which localized in-gap edge states are found, while black areas indicate that all the eigenstates of the system are mostly delocalized.
We also indicate in Fig.~\ref{fig:Heatmap minPR epsilon vs Lx} the value of the Zak phase \eqref{eq:Zak phase}, with the topological phase transitions marked by white dotted lines.

A complex phase diagram appears in Fig.~\ref{fig:Heatmap minPR epsilon vs Lx}, with three distinct topological phase transitions.
The first one is visible for any cavity height, at $\epsilon=0$.
It is the usual transition already present in the original SSH model, in which edge states appear when $d_1 > d_2$ ($\epsilon>0$).
Two additional transitions, one visible for $2.2 \lesssim L_x/a \lesssim 2.8$, and the other for $3 \lesssim L_x/a \lesssim 4.8$, are however solely induced by the longitudinal cavity mirrors.

Considering the transition for $2.2 \lesssim L_x/a \lesssim 2.8$ in Fig.~\ref{fig:Heatmap minPR epsilon vs Lx}, we observe the accordance of the system with the BEC, with, for $\epsilon>0$ ($\epsilon<0$), the appearance (disappearance) of in-gap edge states as the cavity height $L_x/a$ is decreased.
We note that the slight nonaccordance around $\epsilon \sim 0$ arises from finite size effects.
Moreover, a precise study of such a transition (not shown) reveals the closing of the band gap at the edges of the first Brillouin zone, $q=\pm \pi/d$, as for the transition of the original SSH model at $\epsilon=0$.

By now examining the transition for $3 \lesssim L_x/a \lesssim 4.8$, we observe that the presence of localized edge states does not always match the value of the Zak phase. 
Indeed, depending on the values of $\epsilon$ and $L_x$, either the situation discussed in Fig.~\ref{fig:Dispersion triple plot TPT with BEC} or the one discussed in Fig.~\ref{fig:Dispersion triple plot TPT without BEC} may happen, the two latter cases being highlighted in the phase diagram of Fig.~\ref{fig:Heatmap minPR epsilon vs Lx} by blue and red arrows, respectively.
In particular, the black areas visible in regions of nontrivial Zak phase (namely, the bottom center and upper right regions of Fig.~\ref{fig:Heatmap minPR epsilon vs Lx}) correspond to the situation of Fig.~\ref{fig:Dispersion triple plot TPT without BEC}, with edge states that have merged into the upper bulk band.
From the absence of polaritons in the latter bulk band, in-gap edge states just disappear. We note that for larger values of $L_x/a$, as the ones considered in Sec.~\ref{sec:Edge states}, such merging into the bulk would be accompanied with the emergence of polaritonic edge states, showing intermediate values of the participation ratio.

Interestingly, we note in the upper central region of the phase diagram of Fig.~\ref{fig:Heatmap minPR epsilon vs Lx} the presence of two yellow lines, indicating other localized states.
Importantly, these localized states do not correspond to in-gap edge states linked to the topology of the system.
These Tamm-like edge states appear precisely when one of the two bands becomes nearly flat, as can be seen in Fig.~\ref{fig:Dispersion triple plot TPT with BEC}(c).
In contrast to the usual Tamm states we studied in Sec.~\ref{sec:Tamm}, they are here present even in the absence of any energy defect at the ends of the chain.
The same unusual Tamm-like edge states have already been encountered in honeycomb geometries \cite{Plotnik2014,Pantaleon2018} and two-leg ladders \cite{Downing2024}.

\begin{figure}[t]
 \includegraphics[width=\linewidth]{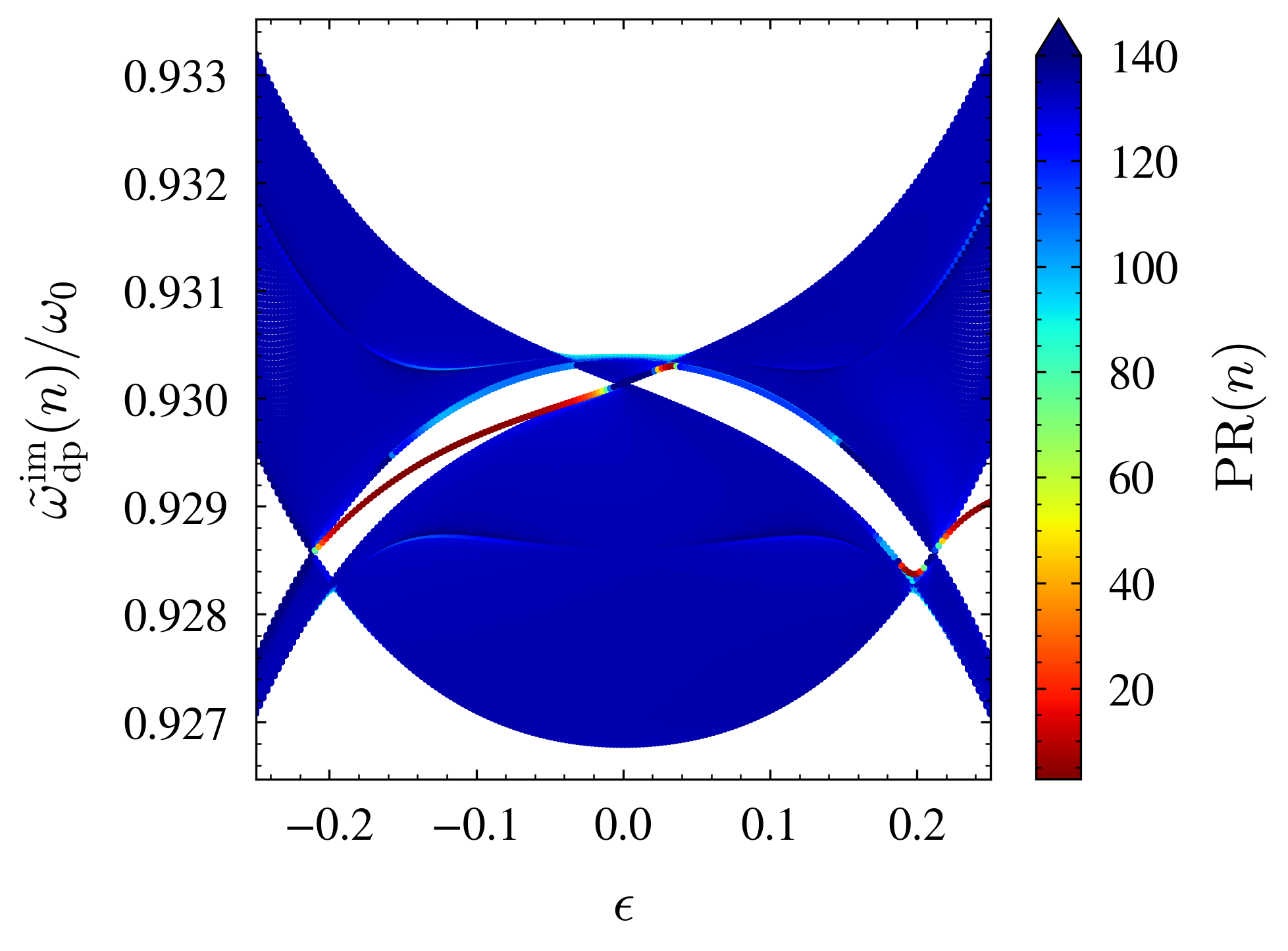}
 \caption{Eigenfrequencies $\tilde{\omega}^\mathrm{im}_\mathrm{dp}(n)/\omega_0$ as a function of the dimerization parameter $\epsilon$, revealing three consecutive gap closings as $\epsilon$ is varied from $-0.25$ to $+0.25$.
 The color code indicates the participation ratio $\mathrm{PR}(n)$ of a given eigenstate $n$.
 In the figure, the cavity height $L_x/a=3.3$, the chain is composed of $\mathcal{N}=100$ dimers, and the distance to the cavity $d_\mathrm{cav}=100d$ is large enough to prevent any effect from the transverse mirrors.}
 \label{fig:Spectrum vs epsilon with PR}
\end{figure}

To conclude this investigation on the effects of the longitudinal cavity mirrors on the system, we present in Fig.~\ref{fig:Spectrum vs epsilon with PR} the eigenspectrum $\tilde{\omega}^\mathrm{im}_\mathrm{dp}(n)$ as a function of the dimerization $\epsilon$ for a cavity height $L_x/a=3.3$.
For such a cavity height, we deduce from the phase diagram of Fig.~\ref{fig:Heatmap minPR epsilon vs Lx} that tuning the dimerization $\epsilon$ allows three distinct topological phase transitions, with the appearance, disappearance, and reappearance of in-gap edge states as $\epsilon$ goes from $-0.25$ to $+0.25$.
The latter transitions are clearly visible in Fig.~\ref{fig:Spectrum vs epsilon with PR}, in which edge states appear in red, the color code representing the participation ratio of each eigenstate $n$, from red (localized) to blue (delocalized).
Moreover, one also observes in Fig.~\ref{fig:Spectrum vs epsilon with PR} the unusual Tamm-like edge states around $\epsilon=0.03$, when the upper band is flat, and around $\epsilon=0.20$, when the lower band is flat.

\section{Conclusions}
\label{sec:Conclusion}

To sum up, we considered a system of dipolar emitters in an optical cavity and focused our attention on physical effects that are often disregarded and which originate solely from the boundary conditions imposed by the cavity mirrors.
In order to decipher between mirror-induced and polaritonic effects, we considered a well-controlled and thoroughly studied model of a chain of emitters embedded in a cuboidal cavity with perfect metallic mirrors \cite{Downing2019,Downing2021,Allard2023}, a system recently realized experimentally \cite{Zhao2024_arXiv}.

We started our investigation by focusing on the influence of the cavity mirrors that are transverse to the direction of the chain.
The position of these mirrors with respect to the chain of emitters, specified in our model by the distance $d_\mathrm{cav}$ (see Fig.~\ref{fig:Sketch}), defines the shape of the electromagnetic modes inside the cavity.
We demonstrated that by placing the transverse mirrors close to the first and last emitters of the chain, the vanishing of the electromagnetic field on the mirror surfaces induces a decrease of the light-matter interaction at the two ends of the chain as compared with its bulk.

Considering first a regularly spaced chain, with constant distances $d_1=d_2$ between all the emitters, we unveiled that such a reduced light-matter coupling at the two ends of the chain leads to the emergence of Tamm edge states, which were initially thought to be of polaritonic origin \cite{Downing2021}.
We then investigated the case of a dimerized chain, with alternating distances $d_1 \neq d_2$ between emitters.
In that case, the system becomes analogous to the SSH model of one-dimensional topological physics.
In the strong light-matter coupling regime, it has been previously shown that the original edge states of the SSH model are lost in the bulk.
However, the system hosts polaritonic edge states formed by polaritons inheriting exponential edge localization \cite{Downing2019,Allard2023}.
Here we demonstrated that the properties of these polaritonic edge states highly depend on the location of the transverse cavity mirrors.
In particular, bringing these mirrors closer to the chain of emitters provides a form of protection of the original edge states against their hybridization into polaritons.

We also examined the specific effects of the mirrors parallel to the direction of the chain.
Through the boundary conditions they impose on the electromagnetic field inside the cavity, such longitudinal mirrors induce image dipoles that renormalize the dipolar degrees of freedom, and hence the eigenspectrum of the system.
Considering again a dimerized chain analogous to the SSH model, we showed that even with highly off-resonant photons with frequencies $\omega_{n_z}^\mathrm{ph} \gg \omega_0$, such renormalizations caused by longitudinal mirrors result in topological phase transitions.
A rich phase diagram has been found, with band-gap closures associated or not with the appearance and disappearance of in-gap edge states in the system.

By demonstrating the possible importance of mirrors independently of the strong light-matter coupling regime, our study should motivate future theoretical works in the polaritonics community to include these often neglected effects.
An important extension of our analysis would be the complexification of the simple model we considered.
Indeed, the addition of multipolar interactions, hence going beyond our description of ideal dipolar emitters, as well as the consideration of nonperfect mirrors may lead to some additional interesting mirror-induced phenomena.

\begin{acknowledgments}
We are grateful to Denis Basko and Clément Tauber 
for insightful discussions that initiated this work.
We thank Charles A.\ Downing for discussions and for his careful reading of our paper. We acknowledge David Hagenmüller for useful comments.
This work of the Interdisciplinary Thematic Institute QMat, as part of the ITI 2021-2028 program of the University of Strasbourg, CNRS, and Inserm, was supported by IdEx Unistra (ANR 10 IDEX 0002), and by SFRI STRAT’US Projects No.\ ANR-20-SFRI-0012 and No.\ ANR-17-EURE-0024 under the framework of the French Investments for the Future Program.
\end{acknowledgments}

\appendix

\section{Dipolar Hamiltonian and image dipoles}
\label{sec:Appendix images}

In this Appendix, we provide details about our modeling, within the standard Coulomb gauge, of dipolar emitters leading to the quasistatic dipolar Hamiltonian \eqref{eq:H_dp with images}.
We start by presenting the Hamiltonian of an ensemble of dipolar emitters in vacuum, and then derive its important modifications induced by the metallic cuboidal cavity through image dipoles.

We consider the subwavelength emitters to be all polarized in the same direction $x$ and arranged on a one-dimensional bipartite lattice along the $z$ direction (see Fig.~\ref{fig:Sketch}).
Each emitter, belonging to the unit cell $m=1,\ldots,\mathcal{N}$ and the sublattice $s=A,B$, has an effective mass $M$ and an effective charge $-Q$.
Its sole dynamical degree of freedom is its displacement field $\mathbf{h}_m^s = h_m^s \hat{x}$, leading to an electric dipole moment $\mathbf{p}_m^s = -Q h_m^s \hat{x}$, oscillating at a resonance frequency $\omega_0 = \sqrt{ Q^2 / Ma^3 }$, where $a$ is the typical dipole length scale.
The above point dipole approximation requires the resonance frequency $\omega_0$ and the dipole length scale $a$ to be in the regime $\omega_0 a /c \ll 1$, where $c$ is the speed of light in vacuum. In the entire paper, we fix $\omega_0 a /c = 0.1$.

The Hamiltonian of such $2\mathcal{N}$-coupled emitters reads
\begin{equation}
    H_\mathrm{dp} = \sum_{s=A,B} \sum_{m=1}^{\mathcal{N}} \left( \frac{  {\mathbf{\Pi}_m^s}^2 }{2M} + \frac{ M\omega_0^2 {\mathbf{h}^s_m}^2}{2} \right) + V_\mathrm{Coulomb}^\mathrm{dip\textrm{-}dip} ,
\label{eq:app:H_dp}
\end{equation}
with $\mathbf{\Pi}_m^s=\Pi_m^s\hat{x}$ the conjugate momentum to $\mathbf{h}_m^s$, and
where the dipole-dipole quasistatic Coulomb coupling writes
\begin{widetext}
\begin{equation}
    V_\mathrm{Coulomb}^\mathrm{dip\textrm{-}dip} = \frac{1}{2} \sum_{s=A,B} \sum_{\substack{m,m'=1\\(m\neq m'\textrm{ $\mathrm{if}$ } s=s')}}^\mathcal{N} \frac{ \mathbf{p}_m^s \cdot \mathbf{p}_{m'}^{s'} - 3\left( \mathbf{p}_m^s \cdot \hat{n}_{m,m'}^{s,s'} \right)\left( \mathbf{p}_{m'}^{s'} \cdot \hat{n}_{m,m'}^{s,s'} \right)}{ |\mathbf{r}_m^s - \mathbf{r}_{m'}^{s'}|^3 }.
\label{eq:app:Coulomb dip-dip}
\end{equation}
\end{widetext}
In Eq.~\eqref{eq:app:Coulomb dip-dip}, the position of an emitter is $\mathbf{r}_m^s = ( L_x/2 , L_y/2 , z_m^s )$, where $z^s_m$ is defined in Eq.~\eqref{eq:zsm}, and the unit vector $\hat{n}_{m,m'}^{s,s'} = ( \mathbf{r}_m^s - \mathbf{r}_{m'}^{s'} ) / | \mathbf{r}_m^s - \mathbf{r}_{m'}^{s'} | $.

We rewrite the dipolar Hamiltonian \eqref{eq:app:H_dp} in the second quantization formalism by introducing the ladder operators
\begin{subequations}
\label{eq:app:bosonic operators}
    \begin{align}
        a_m &= \sqrt{\frac{M\omega_0}{2\hbar}}\,h^A_m + \mathrm{i}\,\sqrt{\frac{1}{2\hbar M \omega_0}}\,\Pi^A_m,\\
        b_m &= \sqrt{\frac{M\omega_0}{2\hbar}}\,h^B_m + \mathrm{i}\,\sqrt{\frac{1}{2\hbar M \omega_0}}\,\Pi^B_m,
    \end{align}
\end{subequations}
which annihilate a dipolar excitation respectively on the $A$ and $B$ sublattice in the unit cell $m$. The above operators obey the bosonic commutation relations $[a_m, {a}_{m'}^\dagger] = [b_m, {b}_{m'}^\dagger] = \delta_{mm'}$.
By adopting the RWA where the number of excitations in the system is conserved, one obtains for Eq.~\eqref{eq:app:H_dp}
\begin{align}
    H_{\mathrm{dp}} =&\; \hbar \omega_0\sum_{m=1}^\mathcal{N} 
    \left(  a_{m}^{\dagger} a_{m}^{\phantom{\dagger}} + b_{m}^{\dagger} b_{m}^{\phantom{\dagger}}\right)
    \nonumber \\ 
    &+ \frac{\hbar\Omega}{2}\sum_{\substack{m,m'=1\\(m\neq m')}}^{\mathcal{N}} f_{m-m'} \left( a_{m^{\phantom{}}}^{\dagger} a_{m^\prime}^{\phantom{\dagger}} + b_{m^{\phantom{}}}^{\dagger} b_{m^\prime}^{\phantom{\dagger}} + \mathrm{H.c.} \right)
    \nonumber \\
    &+ \hbar\Omega\sum_{m,m'=1}^{\mathcal{N}} g_{m-m'} \left( a_{m^{\phantom{}}}^{\dagger} b_{m^\prime}^{\phantom{\dagger}} + \mathrm{H.c.} \right),
\label{eq:app:H_dp quantized}
\end{align}
with $\Omega = (\omega_0/2)(a/d)^3$ the dipolar coupling strength. The  bare intra- and intersublattice couplings read 
\begin{subequations}
\label{eq:app:dipolar couplings}
    \begin{equation}
        f_{m-m'} = \frac{1}{|m-m'|^3}
        \label{eq:app:dipolar coupling f}
    \end{equation}
    and
    \begin{equation}
        g_{m-m'} = \frac{1}{|m-m'-d_1/d|^3},
        \label{eq:app:dipolar coupling g}
    \end{equation}
\end{subequations}
respectively.

So far, the photonic cavity was disregarded. As introduced in the main text, we embed the dipoles inside a cuboidal cavity with perfect metallic plates on all sides, as sketched in Fig.~\ref{fig:Sketch}.
In order for the electric potential to be vanishing on the perfect metallic walls, fictitious, image dipoles located outside of the cavity must be introduced \cite{Jackson2007}.
These additional dipoles couple to the real ones inside of the cavity, inducing an additional Coulomb potential energy, therefore renormalizing the Hamiltonian \eqref{eq:app:H_dp quantized} as \cite{Sturges2020}
\begin{equation}
    H_\mathrm{dp}^\mathrm{im} = H_\mathrm{dp} + V_\mathrm{Coulomb}^\mathrm{dip\textrm{-}im},
\label{eq:app:H_dp,im}
\end{equation}
where the Coulomb potential energy due to image dipoles reads
\begin{widetext}
    \begin{align}
       V_\mathrm{Coulomb}^\mathrm{dip\textrm{-}im} = \sum_{s,s'=A,B} \sum_{m,m'=1}^\mathcal{N}   {\sum_{l,l',l''}}^\prime \frac{  \mathbf{p}_m^s \cdot \mathbf{P}_{m',l,l',l''}^{s'} - 3\left(\mathbf{p}_m^s \cdot \hat{n}_{m,m',l,l',l''}^{s,s'}\right)\left(\mathbf{P}_{m',l,l',l''}^{s'} \cdot \hat{n}_{m,m',l,l',l''}^{s,s'} \right)  }{  2 | \mathbf{r}_m^s - \mathbf{R}_{m',l,l',l''}^{s'}|^3  } .
    \label{eq:app:Coulomb dip-im}
    \end{align}
Here the subscripts $l$, $l'$, and $l''$ label, respectively, all the image dipoles induced by the cavity walls in the $x$, $y$, and $z$ directions, located at positions 
    \begin{equation}
        \mathbf{R}_{m,l,l',l''}^{s} =\left(  [l+1/2]L_x  ,  [l' + 1/2]L_y  , 
         l''L_z + (-1)^{l''} z_m^s - L_z \eta_{l''}  \right),
    \label{eq:app:positions images}
    \end{equation}
\end{widetext}
    where the Boolean function $\eta_{l''} = [(-1)^{l''}-1]/2$, and with dipole moments
    \begin{equation}
    \mathbf{P}_{m',l,l',l''}^{s'} = (-1)^{l'+l''}\left(-Qh_m^s \hat{x}\right).
    \label{eq:app:dipole moments images}
    \end{equation}
In Eq.~\eqref{eq:app:Coulomb dip-im} the unit vector
\begin{equation}
    \hat{n}_{m,m',l,l',l''}^{s,s'} = \frac{  \mathbf{r}_m^s - \mathbf{R}_{m',l,l',l''}^{s'}  }{ | \mathbf{r}_m^s - \mathbf{R}_{m',l,l',l''}^{s'} |  },
\end{equation}
and the primed sum indicates a summation over $(l,l',l'') \in \mathbb{Z}^3 \backslash (0,0,0)$.
We note the factor one-half in Eq.~\eqref{eq:app:Coulomb dip-im} which takes into account that the Coulomb potential energy between real and image dipoles is half of the one between real dipoles \cite{Power1982}.
The alternating signs of the image dipole moments in Eq.~\eqref{eq:app:dipole moments images} originate from the fact that reflections of a dipole polarized onto the $x$ axis into mirrors in the ($x,y$) and ($x,z$) planes are alternately changing direction.

Then, using the quantization scheme presented above, one can rewrite the Coulomb energy \eqref{eq:app:Coulomb dip-im} in terms of the bosonic ladder operators \eqref{eq:app:bosonic operators}.
By considering again the RWA and carefully reorganizing the resulting expression, we finally obtain from the Hamiltonian \eqref{eq:app:H_dp,im} the quasistatic dipolar Hamiltonian within a cavity given in Eq.~\eqref{eq:H_dp with images} of the main text.
In the latter Hamiltonian, the bare dipolar frequencies $\omega_0$ of the emitters, as well as the intra- and intersublattice dipolar couplings \eqref{eq:app:dipolar couplings} are renormalized by the cavity walls into
\begin{subequations}
\label{eq:app:image renormalized quantities}
    \begin{equation}
        \omega_m^{\mathrm{im}, A(B)} = \omega_0 + \delta \omega^{\mathrm{im}, A(B)}_m,
    \label{eq:app:image renormalized frequencies} 
    \end{equation}
    \begin{equation}
        f^{\mathrm{im}, A(B)}_{m,m'} = f_{m-m'} +   \delta f^{\mathrm{im}, A(B)}_{m,m'},
    \label{eq:app:image renormalized f} 
    \end{equation}
    and
    \begin{equation}
       g^\mathrm{im}_{m,m'} = g_{m-m'} + \delta g^\mathrm{im}_{m,m'}.
    \label{eq:app:image renormalized g} 
    \end{equation}
\end{subequations}
The renormalizations read
\begin{subequations}
\label{eq:app:image renormalizations}
    \begin{align}
        \delta \omega^{\mathrm{im}, A(B)}_m =&\; \frac{\omega_0 a^3}{2}{\sum_{l,l',l''}}^{\prime} (-1)^{l'+l''}\nonumber\\
        &\times {I}\!\left( l L_x , l' L_y , l'' L_z - {Z}_{m,m,l''}^{A(B)}\right),
    \end{align}
    \begin{align}
       \delta f^{\mathrm{im},A(B)}_{m,m'} =&\; d^3 {\sum_{l,l',l''}}^{\prime} (-1)^{l'+l''} \nonumber\\
        &\times{I}\!\left( l L_x , l' L_y , l'' L_z - {Z}_{m,m',l''}^{A(B)} \right),
    \end{align}
    and
    \begin{align}
       \delta g_{m,m'}^{\mathrm{im}} =&\; d^3 {\sum_{l,l',l''}}^{\prime} (-1)^{l'+l''}\nonumber\\
        &\times {I}\!\left( l L_x , l' L_y , l'' L_z - {Z}_{m,m',l''}^{AB} \right).
    \end{align}
\end{subequations}
Here we defined the function \cite{Sturges2020}
\begin{equation}
    {I}(x,y,z) = \frac{-2x^2 + y^2 + z^2}{\left(x^2+y^2+z^2\right)^{5/2}},
\label{eq:app:image function}
\end{equation}
as well as the distances
\begin{align}
    Z^{A}_{m,m',l''} &= md - (-1)^{l''}m'd + (\mathcal{N}d+d+d_1) \eta_{l''},\\
    Z^{B}_{m,m',l''} &= md - (-1)^{l''}m'd + (\mathcal{N}d+d-d_1) \eta_{l''},\\
    Z^{AB}_{m,m',l''} &= md - (-1)^{l''}m'd - d_1 + (\mathcal{N}d+d-d_1) \eta_{l''}.
\end{align}

The main effect of the image dipoles on the bare dipolar frequencies $\omega_0$ is a slight redshift, increasing as the cavity dimensions become smaller in units of $a$.
Their effect on the dipolar couplings \eqref{eq:app:image renormalized f} and \eqref{eq:app:image renormalized g} is an exponential suppression of the all-to-all interaction as the cavity dimensions become comparable to the interdipole distances.

In order to numerically compute the expressions \eqref{eq:app:image renormalizations}, one must in principle sum over all the $(l,l',l'') \in \mathbb{Z}^3 \backslash (0,0,0)$, which is not numerically feasible.
In this work, we consider a summation over at least the first $20$ reflections of each dipole in each of the six mirrors constituting the cavity, and truncate the all-to-all dipole interaction to the first $10$ neighboring couplings, in order to better account for its exponential suppression.
We verified that evaluating numerically the quantities \eqref{eq:app:image renormalizations} more accurately do not lead to any qualitative change in the results presented in the paper.

\section{Schrieffer-Wolff transformation}
\label{sec:Appendix Schrieffer-Wolff}

To derive an effective dipolar Hamiltonian, we integrate out the photonic degrees of freedom of the full polaritonic Hamiltonian \eqref{eq:total Hamiltonian} by performing the Schrieffer-Wolff unitary transformation \cite{Schrieffer1966}
\begin{equation}
    \tilde{H}_\mathrm{pol} = \mathrm{e}^S H_\mathrm{pol} \mathrm{e}^{-S} \simeq H_\mathrm{pol} + [S , H_\mathrm{pol}] + \frac{1}{2}[ S , [ S , H_\mathrm{pol} ] ].
    \label{eq:app:SW transformation}
\end{equation}
To eliminate coupling terms of the order of $\Omega\left(\xi^{A/B}_{m,n_z}\right)^2/\omega_0^3$, we identify the anti-Hermitian operator $S$ such that
\begin{equation}
    [S , H_\mathrm{dp}^\mathrm{im}\left(\Omega=0\right) + H_\mathrm{ph}] = - H_{\mathrm{dp}\textrm{-}\mathrm{ph}},
\label{eq:app:SW condition}
\end{equation}
where the Hamiltonians $H_\mathrm{dp}^\mathrm{im}$, $H_\mathrm{ph}$, and $H_{\mathrm{dp}\textrm{-}\mathrm{ph}}$ are given, respectively, in Eqs.~\eqref{eq:H_dp with images}, \eqref{eq:H_ph}, and \eqref{eq:H_dp-ph}, and where we used the fact that the quasistatic dipole-dipole coupling strength $\Omega/\omega_0 \ll 1$.

From the above condition \eqref{eq:app:SW condition}, we find
\begin{align}
    S =&\; -\mathrm{i} \sum_{m=1}^\mathcal{N} \sum_{n_z = 1}^{N_z} \left( \frac{\xi_{m,n_z}^A}{\omega_{n_z}^\mathrm{ph} - \omega_m^{\mathrm{im}, A}} c_{n_z} a_m^\dagger \right. \nonumber \\
    &\left. \; +\; \frac{\xi_{m,n_z}^B}{\omega_{n_z}^\mathrm{ph} - \omega_m^{\mathrm{im},B}} c_{n_z}b_m^\dagger + \mathrm{H.c.} \right).
\label{eq:app:SW S operator}
\end{align}
The dipolar and photonic subspaces are then decoupled to second order in the light-matter coupling strength, with
\begin{equation}
    \tilde{H}_\mathrm{pol} \simeq {H}_\mathrm{dp}^\mathrm{im} + {H}_\mathrm{ph} + \frac{1}{2}[S,H_{\mathrm{dp}\textrm{-}\mathrm{ph}}] \equiv \tilde{H}_\mathrm{dp}^\mathrm{im} + \tilde{H}_\mathrm{ph}.
    \label{eq:app:SW effective Hamiltonian}
\end{equation}
Computing the commutator in Eq.~\eqref{eq:app:SW effective Hamiltonian} and focusing on the dipolar subspace, we obtain the effective Hamiltonian \eqref{eq:SW H_dp with images}.

\section{Fourier transformation}
\label{sec:Appendix Fourier transformation}

Here we detail the derivation and expressions of the Fourier transformed Hamiltonians used in Secs.~\ref{sec:Model} and \ref{sec:Images}.
In order to move in Fourier space, we consider the thermodynamic limit of an infinitely long chain of emitters, with $\mathcal{N} \to \infty$, commensurate with an infinitely long cavity, with length $L_z \sim \mathcal{N}d \to \infty$.
Then, translation invariance allows us to consider periodic boundary conditions in the $z$ direction.

This Appendix is organized as follows. First, we derive the dipolar Hamiltonian in Fourier space including the contributions of image dipoles originating from cavity walls in the $x$ and $y$ directions.
Second, we present the Fourier counterpart of the full polaritonic Hamiltonian \eqref{eq:total Hamiltonian}, of which we show the eigenspectrum in Fig.~\ref{fig:Dispersion Polariton}.
Third, we derive the Fourier counterpart of the effective Hamiltonian \eqref{eq:SW H_dp with images}, by considering an infinitely long chain of emitters in an infinitely elongated cavity.

\subsection{Dipolar Hamiltonian and image dipoles}

We start by rewriting the dipolar Hamiltonian \eqref{eq:H_dp with images} including image renormalizations assuming an infinitely long cavity, namely, a distance $d_\mathrm{cav} \to \infty$, so that we get rid of the cavity walls in the $z$ direction.
Then, we move into wave-vector space through the Fourier transforms\footnote{Note that we consider here the widely used convention of Fourier transforms acting on the cell index only, so that the resulting Fourier Hamiltonian is periodic in the Brillouin zone.}
\begin{subequations}
\label{eq:app:fourier transforms}
    \begin{align}
        a_m &= \frac{1}{\sqrt{\mathcal{N}}} \sum_q \mathrm{e}^{\mathrm{i}mqd}\,a_q,\\
        b_m &= \frac{1}{\sqrt{\mathcal{N}}} \sum_q \mathrm{e}^{\mathrm{i}mqd}\,b_q,
    \end{align}
\end{subequations}
where the bosonic ladder operators $a_q^\dagger$ ($b_q^\dagger$) and $a_q$ ($b_q$) create and annihilate, respectively, a dipolar excitation with resonance frequency $\omega_0$ and quasimomentum $q \in [-\pi/d,+\pi/d]$ on the $A$ ($B$) sublattice.

The Hamiltonian \eqref{eq:H_dp with images} then becomes
\begin{align}
    H_\mathrm{dp}^\mathrm{im} =&\; \sum_q \Big[   \hbar\left(\omega_0^\mathrm{im} + \Omega f_q^\mathrm{im} \right) \left( a_q^\dagger a_q + b_q^\dagger b_q \right)  \nonumber \\
    & +\; \hbar\Omega \left( {g_q^\mathrm{im}}^* a_q b_q^\dagger + {g_q^\mathrm{im}} a_q^\dagger b_q \right)   \Big],
\label{eq:app:H_dp fourier}
\end{align}
with the renormalized quantities
    \begin{equation}
        \omega_0^\mathrm{im} = \omega_0 + \frac{\omega_0 a^3}{2} {\sum_{l,l'}}^{\prime} (-1)^{l'} I\!\left(lL_x , l'L_y, 0\right),
    \end{equation}
    \begin{equation}
    \label{eq:f_FT}
        f_q^\mathrm{im} = 2d^3\sum_{m=1}^{\infty} \cos\left(mqd\right) {\sum_{l,l'}} (-1)^{l'} I\!\left(lL_x , l'L_y, md\right),
    \end{equation}
    and
    \begin{align}
    \label{eq:g_FT}
        g_q^\mathrm{im} =&\; d^3 \sum_{l,l'} (-1)^{l'} I\!\left(lL_x , l'L_y, d_1 \right) \nonumber \\
        & + d^3 \sum_{m=1}^{\infty}  {\sum_{l,l'}}  (-1)^{l'} \left[ \mathrm{e}^{-\mathrm{i}mqd} \,I\!\left(lL_x , l'L_y, md-d_1 \right) \right. \nonumber \\
        & \left. +\; \mathrm{e}^{\mathrm{i}mqd} \,I\!\left(lL_x , l'L_y, md+d_1 \right) \right].
    \end{align}

\subsection{Full polaritonic Hamiltonian}

To get a better insight onto the polaritonic model encapsulated in the Hamiltonian \eqref{eq:total Hamiltonian}, we present in Sec.~\ref{sec:Model} its full diagonalization in Fourier space, without using perturbation theory.
This allows us to reveal the genuine polaritonic nature of the normal modes of the system, through the photonic weight indicated as a color code in the band structure of Fig.~\ref{fig:Dispersion Polariton}.

The derivation of such a band structure has been performed first in Ref.~\cite{Downing2019}.
One first writes the photonic Hamiltonian \eqref{eq:H_ph} at the thermodynamic limit mentioned above.
Considering the length of the cavity $L_z \sim \mathcal{N}d \to \infty$, the longitudinal photonic wave vector $k_z$ is then conserved with the quasimomentum $q$ of the dipolar chain, so that the Hamiltonian \eqref{eq:H_ph} rewrites
\begin{equation}
    H_\mathrm{ph} = \sum_{q,u} \omega_{q,u}^\mathrm{ph} c_{q,u}^\dagger c_{q,u},
\label{eq:app:H_ph fourier umklapp}
\end{equation}
with the $2\pi/d$-periodic dispersion
\begin{equation}
    \omega^\mathrm{ph}_{q,u} = c\sqrt{ \left( \frac{\pi}{L_y} \right)^2 + {q_u}^2 },
\label{eq:app:photon dispersion fourier umklapp}
\end{equation}
where $q_u = q - 2\pi u/d$, and with the index $u \in \mathbb{Z}$ labeling the umklapp bands, or so-called diffraction orders.

The light-matter coupling Hamiltonian \eqref{eq:H_dp-ph}, on the other hand, rewrites
\begin{equation}
   \mathrm{H}_\mathrm{dp\textrm{-}ph} = \mathrm{i}\hbar \sum_{q,u} \xi_{q,u} \left[  \left( a_q^\dagger\,\mathrm{e}^{-\mathrm{i} \chi_{q,u} } + b_q^\dagger\, \mathrm{e}^{\mathrm{i} \chi_{q,u}}  \right)c_{q,u}^{\phantom\dagger} - \mathrm{H.c.} \right],
\label{eq:app:H_dp-ph fourier umklapp}
\end{equation}
with the light-matter coupling strength
\begin{equation}
        \xi_{q,u} = \omega_0\,\sqrt{ \frac{2\pi a^3 \omega_0}{dL_xL_y\omega_{q,u}^\mathrm{ph}}  },
\label{eq:app:light-matter coupling strength fourier umklapp}
\end{equation}
and the phase $\chi_{q,u} = q_u d_1/2$.
Importantly, such light-matter coupling Hamiltonian \eqref{eq:app:H_dp-ph fourier umklapp} takes into account the periodic boundary conditions of the cavity, as visible through the exponential function replacing the sine one in Eq.~\eqref{eq:light-matter coupling functions A and B}.

Umklapp bands, or diffraction orders, may be of importance in the context of very small cavity volumes, as investigated in Sec.~\ref{sec:Images}.
In Sec.~\ref{sec:Model}, as our sole purpose is to qualitatively understand the polaritonic Hamiltonian \eqref{eq:total Hamiltonian}, we may restrict ourselves to the $u=0$ band (and thus we omit the $u$ index in the following), as the higher energy ones do not lead to any sizable difference in Fig.~\ref{fig:Dispersion Polariton}.
One then obtains a three-band Hamiltonian in Fourier space, analytically diagonalizable as Eq.~\eqref{eq:total Hamiltonian fourier}, in which the polaritonic dispersion is the solution of a cubic equation and writes \cite{Downing2019}
\begin{equation}
	\omega^\mathrm{pol}_{qj} = \frac{2\left(\omega_0^\mathrm{im} + \Omega f^\mathrm{im}_q\right) + \omega_q^\mathrm{ph}}{3} + \frac{2\Gamma^\mathrm{im}_q}{3}\cos\left( \frac{ \Phi_q^\mathrm{im} + 2\pi s_j }{3} \right),
\label{eq:app:full polaritonic dispersion}
\end{equation}
with the ordering function $s_j = \left\lceil j/2 \right\rceil + (-1)^j$ and $j \in \{1,2,3\}$ the index labeling, respectively, the mostly photonic upper polariton, and the mostly dipolar medium and lower polaritons.
In the above dispersion, the frequency
\begin{equation}
	\Gamma_q^\mathrm{im} = \sqrt{ 3\Omega^2 |g^\mathrm{im}_q|^2 + 6\xi_q^2 + \left( \omega_q^\mathrm{ph} - \omega_0^\mathrm{im} - \Omega f^\mathrm{im}_q \right)^2 }
\label{eq:app:full polaritonic dispersion frequency Gamma_q}
\end{equation}
has been introduced, as well as the angle
\begin{align}
	\Phi_q^\mathrm{im} =&\;  \arccos \biggl( \frac{ 1 }{{\Gamma^\mathrm{im}_q}^3} \biggl\{ 27\Omega |g^\mathrm{im}_q| \xi_q^2 \cos\left(\phi_q^\mathrm{im}+2\chi_q \right) \nonumber \\
	& + \left( \omega_q^\mathrm{ph} - \omega_0^\mathrm{im} - \Omega f^\mathrm{im}_q \right) \nonumber \\
    & \times \left[  9\left( \xi_q^2 - \Omega^2 |g^\mathrm{im}_q|^2 \right) + \left(\omega_q^\mathrm{ph} - \omega_0^\mathrm{im} - \Omega f^\mathrm{im}_q\right)^2  \right] \biggr\} \biggr),
\label{eq:app:full polaritonic dispersion angle Phi_q}
\end{align}
with the phase $\phi_q^\mathrm{im} = \mathrm{arg}\left(g_q^\mathrm{im}\right)$.

The Hopfield operator diagonalizing the Hamiltonian \eqref{eq:total Hamiltonian fourier}
\begin{equation}
	\gamma_{qj} = {A}_{qj}a_q + {B}_{qj}b_q + {C}_{qj}c_q,
\label{eq:app:full polaritonic Hamiltonian Hopfield operator}
\end{equation}
is a linear combination of the dipolar and photonic degrees of freedom.
The modulus squared of the Hopfield coefficients ${A}_{qj}$, ${B}_{qj}$, and ${C}_{qj}$, normalized as $|{A}_{qj}|^2 + |{B}_{qj}|^2 + |{C}_{qj}|^2 = 1$, represent respectively the part of the polaritonic eigenmode that arises from the dipolar excitation on the $A$ sublattice, from the dipolar excitation on the $B$ sublattice, and from the cavity photon excitation.
Their expressions, which can be extracted from the diagonalization procedure, read
\begin{equation}
	{A}_{qj} = \frac{1}{ \sqrt{2 + \xi_q^2 |\Xi_{qj}|^2} },
\label{eq:app:dipolar weights A}
\end{equation}
\begin{align}
	{B}_{qj} =&\; \frac{1}{ \sqrt{2 + \xi_q^2 |\Xi_{qj}|^2} } \nonumber \\
    & \times \frac{ \xi_q^2 - \left(\omega_q^\mathrm{ph} - \omega_{qj}^\mathrm{pol}\right)\left(\omega_0^\mathrm{im} + \Omega f^\mathrm{im}_q - \omega_{qj}^\mathrm{pol}\right) }{ \Omega {g^\mathrm{im}_q}^* \left( \omega_q^\mathrm{ph} - \omega_{qj}^\mathrm{pol} \right) - \xi_q^2\, \mathrm{e}^{2\mathrm{i}\chi_q} },
\label{eq:app:dipolar weights B}
\end{align}
and 
\begin{equation}
	{C}_{qj} = \frac{-\mathrm{i}\,\xi_q\,\mathrm{e}^{-\mathrm{i}\chi_q}}{ \sqrt{2 + \xi_q^2 |\Xi_{qj}|^2} } \;\Xi_{qj},
\label{eq:app:photonic weights}
\end{equation}
with 
\begin{equation}
	\Xi_{qj} = \frac{ \Omega {g^\mathrm{im}_q}^* - \mathrm{e}^{2\mathrm{i}\chi_q}\left( \omega_0^\mathrm{im} + \Omega f^\mathrm{im}_q - \omega_{qj}^\mathrm{pol} \right) }{ \Omega {g^\mathrm{im}_q}^* \left( \omega_q^\mathrm{ph} - \omega_{qj}^\mathrm{pol} \right) - \xi_q^2 \,\mathrm{e}^{2\mathrm{i}\chi_q} }.
\label{eq:app:frequency Xiq}
\end{equation}

We then define the dipolar and photonic parts of a given polaritonic eigenmode from the modulus squared of the above quantities as, respectively,
\begin{equation}
	\mathrm{D}_{qj} = |{A}_{qj}|^2 + |{B}_{qj}|^2
\label{eq:app:dipolar part fourier}
\end{equation}
and
\begin{equation}
	\mathrm{Ph}_{qj} = |{C}_{qj}|^2,
\label{eq:app:photonic part fourier}
\end{equation}
the latter quantity \eqref{eq:app:photonic part fourier} being the one we display as a color code in Fig.~\ref{fig:Dispersion Polariton}.

\subsection{Effective Hamiltonian}

We now derive a Fourier representation of the effective Hamiltonian \eqref{eq:SW H_dp with images}.
We recall that by considering the thermodynamic limit $ L_z \sim \mathcal{N}d \to \infty$, the transverse cavity walls are suppressed so that the system is translationally invariant in the $z$ direction.
Performing the Fourier transforms \eqref{eq:app:fourier transforms}, we obtain the effective two-band Hamiltonian  $\tilde{H}_\mathrm{dp}^\mathrm{im} = \sum_q {{\psi}}_q^\dagger \tilde{\mathcal{H}}_q^{\mathrm{im}\phantom{\dagger}} {{\psi}}_q^{\phantom{\dagger}}$, with the Bloch Hamiltonian
\begin{equation}
    \tilde{\mathcal{H}}^\mathrm{im}_q = \hbar
    \begin{pmatrix}
        \omega_{0}^\mathrm{im} + \Omega \tilde{f}_q^\mathrm{im} & \Omega \tilde{g}_q^\mathrm{im} \\ 
        \Omega {\tilde{g}_q^{\mathrm{im}\, *}} & \omega_{0}^\mathrm{im} + \Omega \tilde{f}_q^\mathrm{im}
    \end{pmatrix}
\label{eq:app:SW Bloch fourier}
\end{equation}
and the spinor creation operator $\psi_q^\dagger = (a_q^\dagger, b_q^\dagger)$.
In Eq.~\eqref{eq:app:SW Bloch fourier}, the renormalized intra- and intersublattice couplings read in Fourier space
\begin{equation}
    \tilde{f}_q^\mathrm{im} = f_q^\mathrm{im} - \frac{1}{\Omega}\, \sum_{u=-\infty}^{+\infty} \frac{\xi^2_{q,u}}{\omega_{q,u}^\mathrm{ph} - \omega_{0}^\mathrm{im}}
    \label{eq:app:lattice sum f SW}
\end{equation}
and
\begin{equation}
    \tilde{g}_q^\mathrm{im} = g_q^\mathrm{im} - \frac{1}{\Omega} \, \sum_{u=-\infty}^{+\infty} \frac{\xi^2_{q,u}\;\mathrm{e}^{-2\mathrm{i}\chi_{q,u}}}{\omega_{q,u}^\mathrm{ph} - \omega_{0}^\mathrm{im}},
    \label{eq:app:lattice sum g SW}
\end{equation}
where $f_q^\mathrm{im}$ and $g_q^\mathrm{im}$ are given in Eqs.~\eqref{eq:f_FT} and \eqref{eq:g_FT}, respectively.

A Bogoliuobov transformation of the effective Bloch Hamiltonian \eqref{eq:app:SW Bloch fourier} then leads  to the band structure
\begin{equation}
    \tilde{\omega}^\mathrm{im}_\mathrm{dp,q\tau} = {\omega}_{0}^\mathrm{im} + \Omega \tilde{f}^\mathrm{im}_q + \tau\, \Omega |\tilde{g}^\mathrm{im}_q|,
\label{eq:app:bandstructure SW}
\end{equation}
where $\tau=+$ ($-$) denotes the high- (low-)energy band.
We carry out a study of the two-band eigenspectrum \eqref{eq:app:bandstructure SW} in Sec.~\ref{sec:Images}, taking into account umklapp bands $u$ from $-100$ to $+100$.
In order to compute the Zak phase \eqref{eq:Zak phase} in Sec.~\ref{sec:Images}, we also deduce from the Bogoliubov transformation the eigenspinors 
\begin{equation}
\label{eq:app:renormalized_spinors}
\ket{ \tilde{\psi}^\mathrm{im}_{q\tau} } = 
\frac{1}{\sqrt{2}}
\begin{pmatrix}
1 \\
\tau\, \mathrm{e}^{\mathrm{i}\tilde{\phi}^\mathrm{im}_q}
\end{pmatrix},
\end{equation}
where the phase $\tilde{\phi}^\mathrm{im}_q = \arg(\tilde{g}^\mathrm{im}_q)$.

\bibliography{Bibliography}

\end{document}